\documentclass[twocolappendix,numberedappendix]{emulateapj}
\usepackage{amsmath}
\usepackage{float}   
\usepackage{color}
\usepackage{rotating}
\usepackage{placeins} 
\usepackage{comment}
\usepackage{epsfig,graphicx}
\usepackage{epstopdf}

\def\etal{{et~al.}}
\def\kms{{\hbox{km s$^{-1}$}}}

\shortauthors{Grasha \etal}
\slugcomment{Accepted for Publication in the Astrophysical Journal}

\begin{document}
\title{Hierarchical Star Formation in Turbulent Media: Evidence from Young Star Clusters}
\author{K. Grasha\altaffilmark{1}, 
B.G. Elmegreen\altaffilmark{2},
D. Calzetti\altaffilmark{1}, 
A. Adamo\altaffilmark{3}, 
A. Aloisi\altaffilmark{4}, 
S.N. Bright\altaffilmark{4},
D.O. Cook\altaffilmark{5}, 
D.A. Dale\altaffilmark{6}, 
M. Fumagalli\altaffilmark{7},
J.S. Gallagher III\altaffilmark{8},
D.A. Gouliermis\altaffilmark{9, 10}, 
E.K. Grebel\altaffilmark{11}, 
L. Kahre\altaffilmark{12},
H. Kim\altaffilmark{13},  
M.R. Krumholz\altaffilmark{14},  
J.C. Lee\altaffilmark{4,15}
M. Messa\altaffilmark{3}, 
J.E. Ryon\altaffilmark{4}, 
L. Ubeda\altaffilmark{4}
}
\altaffiltext{1}{Astronomy Department, University of Massachusetts, Amherst, MA 01003, USA; kgrasha@astro.umass.edu}
\altaffiltext{2}{IBM Research Division, T.J. Watson Research Center, Yorktown Hts., NY}
\altaffiltext{3}{Dept. of Astronomy, The Oskar Klein Centre, Stockholm University, Stockholm, Sweden}
\altaffiltext{4}{Space Telescope Science Institute, Baltimore, MD}
\altaffiltext{5}{California Institute of Technology, 1200 East California Blvd, Pasadena, CA}
\altaffiltext{6}{Dept. of Physics and Astronomy, University of Wyoming, Laramie, WY}
\altaffiltext{7}{Institute for Computational Cosmology and Centre for Extragalactic Astronomy, Department of Physics, Durham University, Durham, United Kingdom}
\altaffiltext{8}{Dept. of Astronomy, University of Wisconsin--Madison, Madison, WI}
\altaffiltext{9}{Zentrum f\"ur Astronomie der Universit\"at Heidelberg, Institut f\"ur Theoretische Astrophysik, Albert-Ueberle-Str.\,2, 69120 Heidelberg, Germany}
\altaffiltext{10}{Max Planck Institute for Astronomy,  K\"{o}nigstuhl\,17, 69117 Heidelberg, Germany}
\altaffiltext{11}{Astronomisches Rechen-Institut, Zentrum f\"ur Astronomie der Universit\"at Heidelberg, M\"onchhofstr.\ 12--14, 69120 Heidelberg, Germany}
\altaffiltext{12}{Dept. of Astronomy, New Mexico State University, Las Cruces, NM}
\altaffiltext{13}{Gemini Observatory, La Serena, Chile}
\altaffiltext{14}{Research School of Astronomy \& Astrophysics, Australian National University, Canberra, ACT 2611, Australia}
\altaffiltext{15}{Visiting Astronomer, Spitzer Science Center, Caltech. Pasadena, CA}

\begin{abstract}
We present an analysis of the positions and ages of young star clusters in eight local galaxies to investigate the connection between the age difference and separation of cluster pairs.  We find that star clusters do not form uniformly but instead are distributed such that the age difference increases with the cluster pair separation to the 0.25--0.6 power, and that the maximum size over which star formation is physically correlated ranges from $\sim$200~pc to $\sim$1~kpc.  The observed trends between age difference and separation suggest that cluster formation is hierarchical both in space and time:  clusters that are close to each other are more similar in age than clusters born further apart.  The temporal correlations between stellar aggregates have slopes that are consistent with turbulence acting as the primary driver of star formation.  The velocity associated with the maximum size is proportional to the galaxy's shear, suggesting that the galactic environment influences the maximum size of the star-forming structures. 
\end{abstract}

\section{Introduction}\label{sec:intro}
Star formation is believed to be hierarchical in space and time as a result of turbulence and self-gravity \citep{scalo85,li05,elmegreen10,kritsuk13} within the interstellar medium (ISM).  The largest scales of the hierarchical structure are star-forming disks and within them are the accumulation of smaller components in the hierarchy: unbound cluster and stellar complexes, star clusters, and individual stars.  Stars mostly form together in some sort of ensemble of clusters and associations \citep{lada03} and it is currently thought that they rarely form in isolation \citep[see, however, ][]{lamb16}.  Furthermore, star clusters are clustered with respect to each other \citep{efremov78} in large complexes, imprinted with the fractal structure of the Giant Molecular Clouds (GMCs) from which they are born \citep{scalo85,elmegreenfalgarone96,sanchez10}, slowly dispersing as the clusters age.  Characterizing the correlation behavior of star formation across a range of galaxy properties provides a validation of this picture and a crucial understanding of how star formation progresses in both space and time across galactic scales.  

Within the framework of star formation models that are regulated by turbulence, gas compression will break larger clouds into successively smaller ones, giving rise to the observed hierarchical structure \citep{elmegreen99,vazquezsemandeni09,hopkins13}.  Such turbulent fragmentation processes, in addition to creating a hierarchy in the distribution of star clusters' properties (e.g., mass, size), will also create correlations with star cluster ages; in a structure where clusters have formed out of the same GMC, clusters that form close together will have closer ages compared to clusters that are further apart.  Coeval cluster pairs have been observed both within the Large Magellanic Cloud \citep[LMC;][]{bhatia88,dieball02} and the Milky Way \citep[MW;][]{desilva15}.  Thus, within this picture, larger structures display older ages.  This is interpreted as the duration of star formation proceeding faster in smaller regions than in larger ones \citep{efremov98}, in proportion to the turbulent crossing time.   

Turbulence-driven star formation predicts that the age of star-forming structures will increase in proportion to the square root of the size \citep{elmegreen96}.  If the age-separation effects were driven simply by the stellar drift of the structure, a linear relation between the structures would be expected \citep{blaauw52, zwicky53}.  
A diffusion-driven expansion would produce a squared relation between age and size, in agreement with expectations from a random walk: the total distance traveled by a random scattering process is related to the square root of the number of random steps taken, where the number of steps is proportional to the time.  These are testable predictions, when the ages of the structures can be determined with sufficient accuracy. 

The age-separation distribution of star clusters was originally investigated within the LMC by \citet{efremov98}, finding that the average age difference between pairs of star clusters increases with their separation as $\Delta t ({\rm Myr}) \sim 3.3 \times S({\rm pc}) ^{0.35 \pm 0.05}$ up to 780 pc.  In a similar venue, \citet{delafuentemarcos09} found local MW cluster pairs to also exhibit age differences with their separation with a $0.40 \pm 0.08$ power once the effects of incompleteness and cluster dissolution are taken into account.  These relationships are similar to the size-linewidth relation of GMCs that show that the crossing time in a GMC increases as the square root of the size of the star forming region \citep{larson81,elmegreen96}.  Numerical simulations investigating the propagation of star formation
in a turbulent medium by \citet{nomura01} derive an age-separation relation to the 0.5 power for star clumps with separations $>$50 pc, comparable with the relation seen in both \citet{efremov98} and \citet{delafuentemarcos09}.  The hierarchy in the star cluster structures is expected to have an upper limit in size; beyond this, separate regions form independently from one another \citep{elmegreen14}, where turbulence can no longer regulate the cluster positions.  Thus, the cluster relation should flatten at the maximum size of a galaxy's star-forming region, resulting in a turnover where the age difference between cluster pairs becomes a random function of the separation.

In this work, we analyze the average age difference between pairs of star clusters in eight local galaxies to investigate their hierarchical distribution, generalizing previous results beyond the MW and LMC.  Our galaxy and cluster sample selection is described in Section \ref{sec:sample}.  The age difference versus separation for cluster pairs is presented in Section \ref{sec:agesep}.  We discuss our results in Section \ref{sec:discussion} and summarize the findings in Section \ref{sec:summary}.

\section{Sample Selection}\label{sec:sample}
The eight galaxies analyzed here were observed as part of the HST Treasury program Legacy ExtraGalactic UV Survey\footnote{https://archive.stsci.edu/prepds/legus/dataproducts-public.html} \citep[LEGUS;][]{calzetti15}, a Cycle 21 survey of 50 nearby star-forming galaxies observed in NUV, U, B, V, and I with WFC3/UVIS.  The galaxy images have all been drizzled to a common scale resolution of the native WFC3 pixel size (0.0396 parsecs per pixel).  The eight galaxies of this study (seven spirals and one irregular dwarf) are those for which star cluster catalogs have been produced by the LEGUS team at the time of this writing.  General descriptions of the LEGUS survey, observations, and data products are available in \citet{calzetti15}.

\subsection{Star Cluster Identification and Selection}\label{sec:cluster}  
A detailed description of the cluster selection, identification, photometry, and SED fitting is presented in \citet{adamo17}.  We summarize here briefly the aspects that are important for the current paper. 

Catalogs of candidate stellar clusters are first identified using SExtractor \citep{bertin96} on a white-light image generated by using the five available photometric bands.  This automated process produces a list of clusters with their position in pixel coordinates, number of filters the source was detected in, and the concentration index (CI; difference in magnitudes measured with an aperture of radius 1 pixel and an aperture of radius 3 pixels) of the source.  The candidate clusters must satisfy two criteria in this automated process: (1) the CI in the V-band must exceed the stellar CI peak value; and (2) photometric error of $\sigma_{\lambda} \leq 0.35$ mag in at least two bands (the reference V band and either B or I band).  Corrections for foreground Galactic extinction \citep{schlafly11} are applied to the photometry.  The concentration index cutoff between stars and star clusters are determined independently for each galaxy and for each pointing.  

Before photometry is performed, the automated cluster candidates are down-selected through excluding any sources that are not detected with a 3$\sigma$ detection in at least four of the five available photometric bands.  The cluster candidates then undergo SED fitting to measure their physical properties (age, color excess E(B--V), and mass) using deterministic stellar population models \citep[Yggdrasil;][]{zackrisson11}.  The conditions placed on the photometry result in uncertainties of the ages and masses of $\sim$0.1 dex.  The cluster SED fitting implement Padova isochrones, solar metallicity, and a range of extinction/attenuation curves.  For this analysis, we select the SED fits performed with the starburst attenuation curve \citep{calzetti00}, which provides reasonable fits for both spiral and dwarf galaxies.  

The sources in the automated catalogs include sources that are not star cluster candidates, such as background galaxies, foreground stars, multiple star pairs in crowded regions, and bad pixels and/or edge effects.  To remove these contaminants, we visually inspect each cluster candidate that has an absolute magnitude brighter than $-6$ mag in the V-band.  Each source is assigned one of four possible classifications: (1) class 1 clusters are compact and symmetrical, displaying a homogeneous color and a FWHM that is more extended than the stars within the galaxy, (2) class 2 clusters are compact and elongated, displaying elliptical light profiles; (3) class 3 clusters are non-compact and reveal multiple peaks on top of diffuse emission; and (4) class 4 sources are contaminants within the catalog; we remove these interlopers from the final {\it visually inspected} cluster catalog.  The final cluster classifications are based on inspection by three separate individuals within the LEGUS team, excluding NGC 1566, where 368 of the clusters (33\% of the total) were identified with only one human classification.  

The clusters in this study comprise clusters with class 1, 2, and 3, which we further down-select to include only clusters with ages less than 300 Myr to ensure that we are not strongly affected by incompleteness due to evolutionary fading (see Section \ref{sec:dissolution}) or cluster dissolution.  Table \ref{tab1} lists the galaxies and number of young cluster candidates ($<300$ Myr) within each galaxy.  

We exclude the clusters within the strong star-forming region in the north-east corner of NGC 5194 to avoid introducing bias from the structure, bringing the total clusters with ages $<$300 Myr from 1171 to 821.  A detailed investigation on the star clusters of NGC 5194 be performed in a forthcoming paper.

\begin{deluxetable*}{lcccccccccc}
\tablecolumns{11}
\tabletypesize{\scriptsize}
\tablecaption{Galaxy Properties\label{tab1}}
\tablewidth{0pt}
\tablehead{
\colhead{Galaxy} 	& 
\colhead{Morph} 	& 
\colhead{Dist.} 		& 
\colhead{R$_{25}$} & 
\colhead{Morph.} 	& 
\colhead{SFR$_{\rm UV}$} & 
\colhead{$N_{\rm clusters}$} & 
\colhead{CI$_{\rm cut}$}	&
\colhead{Scale} 			&
\colhead{Inclination} 	&
\colhead{Position Angle} 					
\\
\colhead{} 					& 
\colhead{} 					& 
\colhead{(Mpc)} 			& 
\colhead{(kpc)}			& 
\colhead{} 					& 
\colhead{(M$_{\odot}$~yr$^{-1}$)} 		& 
\colhead{} 					& 
\colhead{(mag)} 			&
\colhead{(pc/px)} 			&
\colhead{(deg.)} 	&
\colhead{(deg.)} 					
\\
\colhead{(1)} & 
\colhead{(2)} & 
\colhead{(3)} & 
\colhead{(4)} & 
\colhead{(5)} & 
\colhead{(6)} & 
\colhead{(7)} &
\colhead{(8)} &
\colhead{(9)} &
\colhead{(10)} &
\colhead{(11)}
}    
\startdata    
NGC 7793 & SAd		& 3.44 	& 4.65(0.11)	& 7.4(0.6) & 0.52 & 343		& 1.3(e)/1.4(w)	& 0.66 & 47.4 & 90\\ 
NGC 1313 & SBd		& 4.39 	& 5.8(0.3)		& 7.0(0.4) & 1.15 & 673		& 	1.4 					& 0.83 & 40.7 & 39\\
NGC 3738 & Im			& 4.90 	& 1.78(0.08)	& 	9.8(0.7) & 0.07 & 254		& 1.4  					& 0.94 & 40.5 & 141\\
NGC 6503 & SAcd		& 5.27	& 5.44(0.13)	& 	5.8(0.5) & 0.32 & 283		& 1.25 					& 1.1  	& 70.2 & 120\\
NGC 3344 & SABbc	& 7.0 	& 7.23(0.17)	& 	4.0(0.3) & 0.86 & 388		& 1.35 					& 1.3  	& 23.7 & 140\\
NGC 5194 & SAbc		& 7.66 	& 12.5(0.3)		& 4.0(0.3)  & 6.88 & 821	& 	1.35  				& 1.5 	& 51.9 & 163\\ 
NGC 628   & SAc		& 9.9 	& 13.1(0.4)		& 	5.2(0.5) & 3.67 & 1205	& 1.4(c)/1.3(e)  	& 1.9 	& 25.2 & 25\\
NGC 1566 & SABbc	& 13.2	& 15.9(0.8)		&	4.0(0.2) & 5.67 & 1099	& 1.35 					& 2.5	& 37.3 & 32
\enddata
\tablecomments{
Columns list the: 
(1) galaxy name, ordered by increasing distance; 
(2) morphological type as listed in NED, the NASA Extragalactic Database; 
(3) distance from \citet{calzetti15};  
(4) optical radius of the galaxy R$_{25}$ from NED;  
(5) RC3 morphological T--type as listed in Hyperleda (http://leda.univ-lyon1.fr); 
(6) Star Formation Rate (SFR), calculated from the GALEX far-UV, corrected for dust attenuation as described in \citet{lee09}; 
(7) number of star clusters with ages $<300$ Myr; 
(8) concentration index (CI) cutoff between stars and star clusters (see Section \ref{sec:sample}).  NGC 7793 and NGC 628 have different CI cutoffs for each pointing [central (c), east (e), or west (w)], labeled separately; 
(9) pixel resolution in parsec pixel$^{-1}$; 
(10) inclination, in degrees, from \citet{calzetti15}; and
(11) position angle (P.A.), measured in degrees (0 to 180) from the North to the East.  A P.A.$=0$ corresponds to a galaxy with the longest axis oriented along the North-South direction.
Numbers in parentheses indicate $1\sigma$ uncertainties in the final digit(s) of listed quantities, when available. 
}
\end{deluxetable*}


\subsection{Incompleteness and Selection Effects}\label{sec:completeness}
Completeness tests are performed and discussed in \citet{adamo17} on NGC 628 at a distance of 9.9 Mpc; we briefly summarize those tests here.  Completeness tests on the cluster catalogs show that the catalogs are complete down to the CI cutoff of each galaxy, as listed in Table \ref{tab1}, corresponding to clusters with $R_{\rm eff} \sim 1$~pc at 10 Mpc.  With the exception of NGC 1566, NGC 628 is the most distant galaxy in our list (Table \ref{tab1}), thus we expect the other catalogs to be complete to a smaller effective radius.  A CI cut is necessary in order to remove stellar contaminants from the catalog.  Size distributions of star clusters have been shown to display a log-normal distribution that typically peaks around 3 parsecs across galaxies \citep{ryon17}, and hence, we do not expect any bias within the catalogs as the typical cluster radius is well above the detection limit of 1 parsec.  As a result, the distance of the galaxy is not expected to impact the cluster recovery fraction.  Crowding effects between the inner and outer regions of a galaxy is also negligible on the cluster catalogs.  

In addition to using deterministic models to deriving the physical properties of the stellar clusters, the cluster properties are also derived using a Bayesian analysis method to stochastically-sample cluster evolutionary models, as performed by \citet{krumholz15} through implementation of the Stochastically Lighting Up Galaxies \citep[SLUG;][]{dasilva12,krumholz15b} code.  Stochastic effects of the IMF become progressively more important for deriving accurate ages and masses for clusters with masses below $\sim$5000~M$_{\odot}$.  \citet{krumholz15} finds in both NGC 628 and NGC 7793 that the global properties of the cluster populations are relatively similar between the conventional deterministic and stochastic fitting procedures.  Section \ref{sec:sto} shows how our main results are nearly the same when the cluster properties (age and mass) are derived with stochastic models.  As the biases between the two methods (deterministic and stochastic) of deriving the cluster properties averages out for the mean of the entire population \citep{krumholz15}, our results are fairly insensitive to the assumption of incomplete sampling of the IMF at the low-mass cluster range with deterministic fitting.  

The LEGUS cluster catalogs are fundamentally luminosity limited, as seen in the rising top envelope of the age-mass diagram of clusters \citep{grasha15,adamo17}, biasing the catalogs towards younger clusters.  However, as shown in \citet{adamo17} for NGC 628 and NGC 5194, a cut in absolute V magnitude of $-6$ mag is more conservative than the depth of the V-band image, i.e., $-6$ mag is more luminous than the detection limit.  In some parts of this paper (e.g., Section \ref{sec:masscut}) we apply mass cuts to the clusters, as a mass-limited sample will prevent bias in the age distribution caused by a luminosity-limited sample of young clusters.  At 10 Mpc, an age limit of 200 Myr yields a complete cluster sample down to masses of 5000 M$_{\odot}$ \citep{adamo17}, i.e., close to the mass limit where stochastic sampling of the IMF begins to become important.  Our absolute V magnitude limit is determined by the detection limits of the LEGUS sample, which aims at selecting down to $\sim$1000 M$_{\odot}$, 6 Myr old clusters with color excess E(B--V)= 0.25 \citep{calzetti15}. The only exception to the above is NGC 1566, located at a greater distance than NGC 628.  For NGC 1566, a higher  absolute V magnitude cut of $-$8 mag needs to be applied, resulting in a smaller age range for which our catalogs are complete to 5000 M$_{\odot}$.  

In Section \ref{sec:masscut} below, we test for selection effects by repeating our analysis after making mass cuts at both 3000~M$_{\odot}$ and 5000~M$_{\odot}$, which yield complete samples up to ages of 100 Myr and 200 Myr, respectively.  If completeness affects or drives the maximum possible age of a star-forming region, we would expect to see that in the results because the age limit changes as a function of the mass cut \citep[see Figure 3 of][]{grasha15}, and thus, we would expect a correlation between the age to which the sample is complete and the maximum age in the observed $\Delta t - R$ correlation.  In Section \ref{sec:masscut}, we find that the effect of mass cuts on the results is negligible and well within the uncertainties of the data.  Therefore, all analyses throughout this paper include all clusters $<300$ Myr and we do not apply a mass cut in order to strengthen available statistics, as our results are not affected by completeness issues, either in age or mass.

\subsection{Deprojection of the Galactic Disk}\label{sec:deproj}
It is essential to properly account for and deproject the positions of the star clusters to accurately use their spatial distributions free from the effect of projection.  We deproject the pixel coordinate positions of all the stellar clusters in each galaxy by assuming that each galaxy can be well-described with an axisymmetric flat rotating disc.  The deprojection of the cluster positions is performed in a two step process.  

The $x$ and $y$ pixel coordinates are first rotated by the position angle $\alpha$, determined by the orientation of the observed field of view for each galaxy.  The rotated coordinates $x'$ and $y'$ are determined as $x' = x \cos \alpha + y \sin\alpha$ and $y' = -x \sin \alpha + y \cos \alpha$.  The rotated coordinates are then deprojected by the inclination angle $i$ of the galaxy as $x_{\rm deproject} = x'$ and $y_{\rm deproject} = y'/\cos i$.  We use the deprojected coordinate positions $x_{\rm deproject}$ and $y_{\rm deproject}$, converted to a physical scale within each galaxy, for all calculations involving the positions of the clusters using the spatial resolution of a pixel at the distance of each galaxy, as listed in Table \ref{tab1}.  

\section{Analysis and Results}\label{sec:agesep}

\subsection{$\Delta t - R$ Relation}\label{sec:delt}
Following the work of \citet{efremov98}, we consider all clusters younger than 300~Myr within each galaxy, 
and compute the absolute age difference $\Delta t$ between each pair as a function of their deprojected (see Section \ref{sec:deproj}) physical separation (R) for nine equally spaced separation bins in log space.  We allow pair separations up to 8000 parsecs in each galaxy, performing a least-squares fit to the data following a double power-law as, 
\begin{equation}\label{eq:1}
\Delta t ({\rm Myr}) =  \left\{
  \begin{array}{lr}
  A_1 \times R({\rm pc}) ^{\alpha} & R \leq R_{\rm max} \\ 
  A_2 \times R^0 & R > R_{\rm max}.
  \end{array} 
\right.
\end{equation}
with $\alpha$ the slope before the breakpoint at a maximum separation $R_{\rm max}$ where the age--separation trends flatten and $A_1$ and $A_2$ the intercept before and after the breakpoint, respectively.   We force the slope at scale lengths larger than $R_{\rm max}$ to zero and the location of $R_{\rm max}$ is a free parameter in the fit, determined where chi-squared is minimized.  We take into account and propagate the uncertainties of individual cluster ages for each cluster pair for the $\Delta t$ calculation when making the mean age-difference at each bin.  The number of bins are chosen as a compromise between available statistics, where we require a minimum of 10 data points in each bin, and resolution, though on average, the smallest bin for each galaxy has over 20 pairs.  The influence of the choice of bin size is investigated further in Section \ref{sec:agedependency}.  

Assuming that cluster pairs are members of the same region, the observed turnover is a measure of a galaxy's maximum star-forming region size, and the age differences at the same separation can be used as proxies for the duration or lifetime of star formation within the region.  Table \ref{tab2} lists the power-law fit parameters for each galaxy. 

\begin{deluxetable*}{lcccccc}
\tablecolumns{7}
\tabletypesize{\scriptsize}
\tablecaption{Age Difference and Spatial Separation Results \label{tab2}}
\tablewidth{0pt}
\tablehead{
\colhead{Galaxy} 	& 
\colhead{Intercept} & 
\colhead{Slope} 		&
\colhead{Max Age} &
\colhead{$R_{\rm max}$}&
\colhead{Velocity}	&
\colhead{$v_{\rm S}$}	
\\
\colhead{} 					& 
\colhead{($A_1$)} 		& 
\colhead{($\alpha$)} 	&
\colhead{(Myr)} 			& 
\colhead{(pc)} 			&
\colhead{(\kms)} &
\colhead{(\kms)} 
\\
\colhead{(1)} & 
\colhead{(2)} & 
\colhead{(3)} & 
\colhead{(4)} & 
\colhead{(5)} & 
\colhead{(6)} & 
\colhead{(7)} 
}    
\startdata            
NGC 7793 & 4.0(0.3)	& 0.47(0.06)	& 48(19)	& 203(30)	& 4.0(1.1)	& 2.0\tablenotemark{1,2} \\ 
NGC 1313 & 16(1)		& 0.26(0.08)	& 85(26)	& 585(183)	& 6.1(1.4)	& 2.4\tablenotemark{3} \\
NGC 3738 & 9.0(0.5)	& 0.24(0.04)	& 45(9)	& 869(152)	& 15(2) 		& 5.0\tablenotemark{4} \\
NGC 6503 & 1.7(0.3)	& 0.6(0.2)		& 62(21)	& 275(104)	& 5.7(2.5) 	& 1.9\tablenotemark{5} \\
NGC 3344 & 1.0(0.2)	& 0.6(0.2)		& 41(14)	& 338(131)	& 8.5(3.1) 	& 3.7\tablenotemark{6} \\
NGC 5194 & 6.8(1.0)	& 0.36(0.07)	& 83(20)	& 947(231)	& 13(3) 		& 5.5\tablenotemark{7} \\
NGC 628   & 7.4(0.6)	& 0.33(0.07)	& 66(18)	& 788(179)	& 13(4) 		& 6.9\tablenotemark{7,8} \\
NGC 1566 & 2.3(0.2)	& 0.41(0.14)	& 30(9) 	& 508(179)	& 20(7) 		& 10\tablenotemark{9} 
\enddata
\tablecomments{
Power-law fits for the equally spaced separation bins in log space for the galaxies in Figure \ref{fig:agespread}.  Columns list the: 
(1) galaxy name, ordered by increasing distance; 
(2) intercept $A_1$ from Eq. \ref{eq:1};
(3) slope $\alpha$ from Eq. \ref{eq:1}; 
(4) $A_2$, the average age difference between cluster pairs at the turnover;  
(5) size $R_{\rm max}$ of the star forming region at the turnover; 
(6) traveling velocity (size/average age) of the region at the turnover; and
(7) average shear velocity between cluster pairs at $R_{\rm max}$ and references for the rotation curves used to derive the shear.  
Numbers in parentheses indicate $1\sigma$ uncertainties in the final digit(s) of listed quantities, when available. \\
References for rotation curves: 
(1) -- \citet{carignan90}; 
(2) -- \citet{dicaire08}; 
(3) -- \citet{ryder95}; 
(4) -- \citet{oh15}; 
(5) -- \citet{greisen09}; 
(6) -- \citet{verdes00}; 
(7) -- \citet{daigle06}; 
(8) -- \citet{combes97}; 
(9) -- \citet{aguero04}.  
}
\end{deluxetable*}

\begin{figure*}
\includegraphics[scale=0.44]{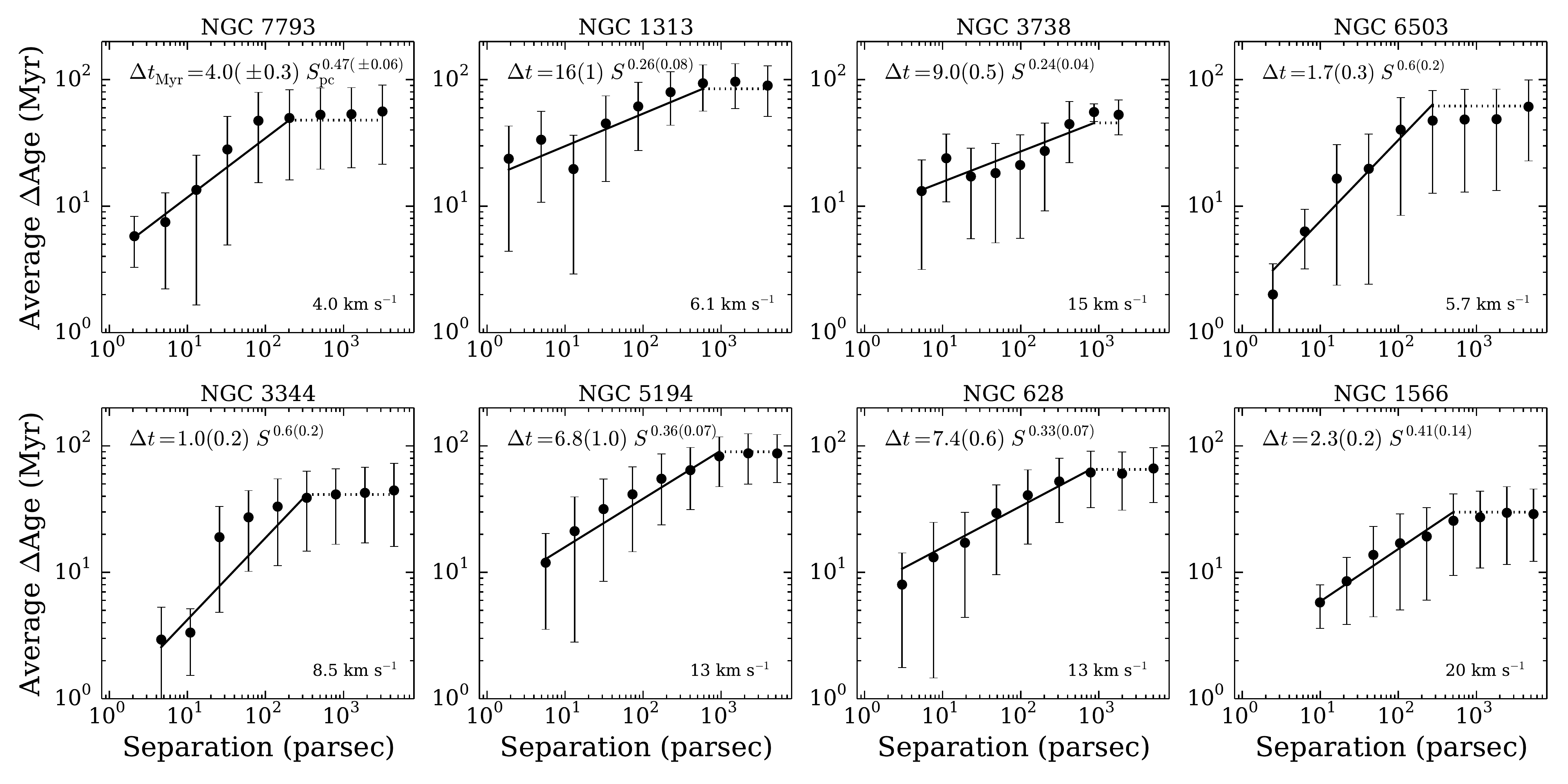}
\caption{
The age difference between cluster pairs as a function of separation between the cluster pairs.  The black data points and error bars show the average age and $1\sigma$ spread for each cluster pair in each separation bin.  The bins are logarithmically-spaced, which provide optimal sampling (see Section \ref{sec:agedependency}).  The average age difference between cluster pairs increases systematically with their separation, indicating that the duration of star formation is longer for larger regions and that younger star-forming structures are less extended than older regions.  A linear fit to the $\Delta t - R$ relation is shown in the solid black line and the dotted line shows where the relationship is flat beyond the breakpoint $R_{\rm max}$.  The lower right hand corner of each panel lists the average velocity at the breakpoint in the double power law.  
\label{fig:agespread}}
\end{figure*}

Figure \ref{fig:agespread} shows the average age-separation as a function of increasing separation between cluster pairs in addition to the power-law fit to each galaxy (eq. \ref{eq:1}).  The maximum separation of correlated cluster pairs corresponds to sizes $\sim$200--1000~pc and average age separations of $\sim$20--100~Myr. 

From Figure \ref{fig:agespread}, we calculate a velocity from the ratio of the size of the turnover $R_{\rm max}$ to the average age difference at that size.  This velocity is the average speed at which turbulence moves through the star-forming region.  The velocity is listed in the lower right hand corner of each panel in Figure \ref{fig:agespread}.

\subsection{Impact of Mass Cuts on the $\Delta t - R$ Relation}\label{sec:masscut}
\begin{figure}
\includegraphics[scale=0.42]{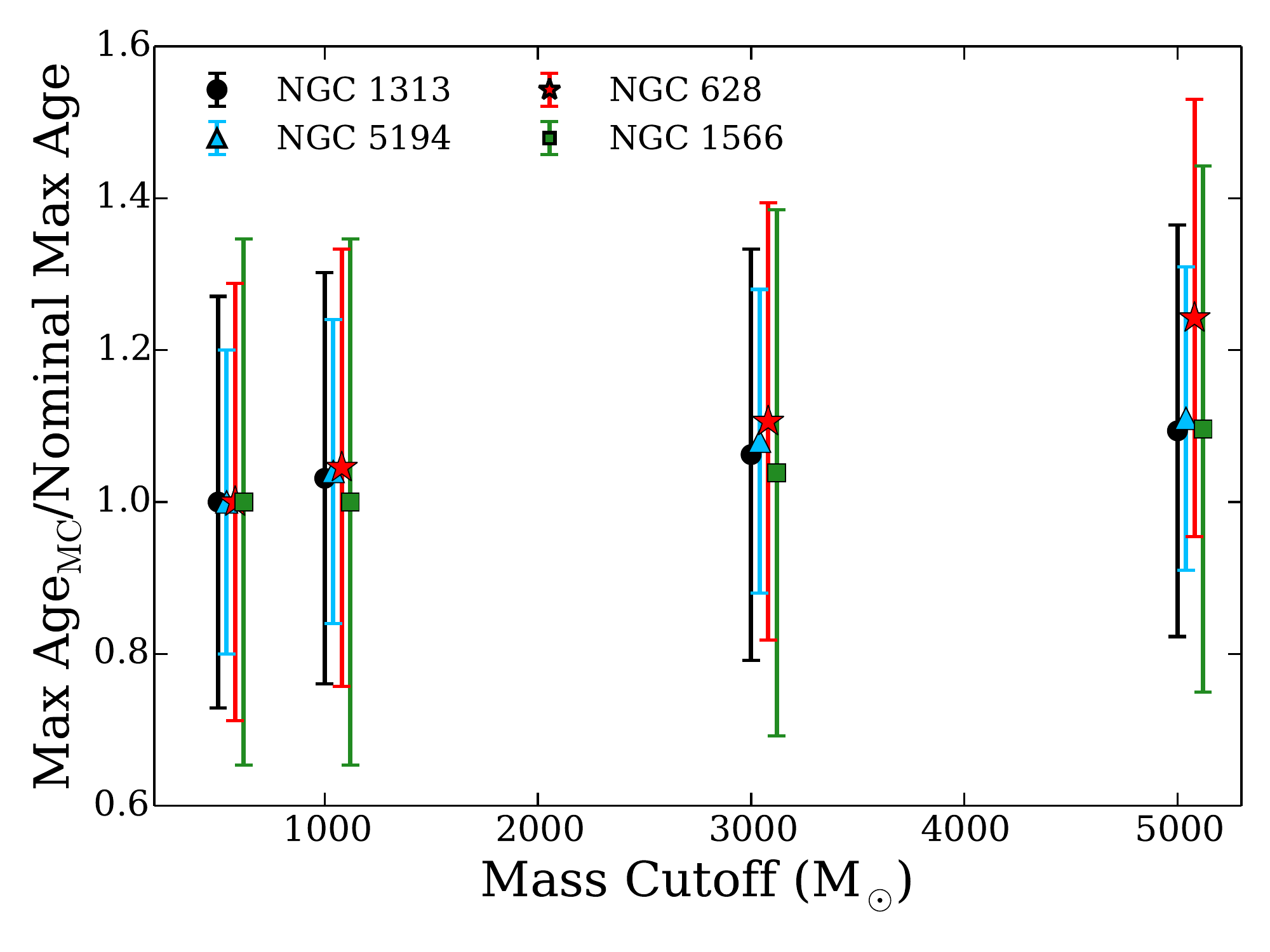}
\caption{
The maximum age for each galaxy at different mass cutoff limits normalized by the nominal maximum age, derived for a minimum mass limit of 500 M$_{\odot}$ (Table \ref{tab2}) as a function of the cutoff mass at 1000, 3000, and 5000 M$_{\odot}$.  If the $\Delta t - R$ relation is a result of the luminosity limit of the sample we would expect an increase in the maximum age as we increase the mass cutoff.  There is a slight upward trend, but the values are well within the errors, implying that the catalog completeness is not responsible for driving the $\Delta t - R$ relation.  The points corresponding to different galaxies are shifted slightly along the x-axis as to not lie on top of each other.  
\label{fig:masscut}}
\end{figure}

If completeness in age and/or mass drives the observed separation of cluster pairs, we would expect to observe a correlation between the limiting age we are sensitive to at a given mass cut and the maximum age at the turn over point.  We repeat the analysis in Section \ref{sec:delt} after applying different mass cuts for the galaxies with enough statistics to do a proper study: NGC 1313, NGC 5194, NGC 628, and NGC 1566 to see if we see a trend for an increase in the age accompanying increases in the mass cut.  Figure \ref{fig:masscut} shows negligible increase in the maximum age of the $\Delta t - R$ relation, about 10\% or less, which is well within the errors.  We do note that NGC 628 does show a higher increase in $R_{\rm max}$ at the largest mass cut, about 25\%.  However, this is still within the error bars, and, as we show in Section \ref{sec:tests}, the $\Delta t - R$ correlation for NGC 628 disappears when we randomize the clusters positions and ages, demonstrating that the correlation is not an effect of selection biases.

Additionally, for all the galaxies in Table \ref{tab2}, the maximum age is always shorter than the completeness age limit of 200 Myr at the mass cut-off of 5000~M$_{\odot}$.  We conclude that the luminosity-limited nature of our catalog, and therefore, potential biases induced by this on the ages, are not driving the observed $\Delta t - R$ correlation.  

\subsection{Shear and Global Galactic Properties}\label{sec:shear}
To better understand the impact of galactic shear on the turbulent velocities, we remove the effect of shear in all galaxies at the turnover separation $R_{\rm max}$ between all clusters in all galaxies.  First, we calculate the angular velocity as $\Omega = V_{\rm rot}/R_{\rm g}$, where $V_{\rm rot}$ is the rotational velocity at the galactocentric position $R_{\rm g}$ of each cluster.  Then we take the average galactocentric distance $R_{\rm g}$ multiplied by the difference in angular rates, $\Delta \Omega$, for each pair of clusters within the $R_{\rm max}$ separation bin.  This is the relative velocity from shear for each pair, $V_{\rm S}$, which we subtract from the turbulent velocity to get a velocity that is corrected for the effect of shear.  The relative velocity from shear for each galaxy is listed in Table \ref{tab2} along with the references from which the rotation curves are taken.  $V_{\rm S}$ quantifies the average difference in azimuthal velocities on the scale of the largest separation and informs on how much shear influences the relative velocity at the edge of a star-forming region.  We find that in general the contribution of shear to the measured velocity between cluster pairs is at most only slightly greater than the 1$\sigma$ error of the velocity. 

\begin{figure}
\includegraphics[scale=0.42]{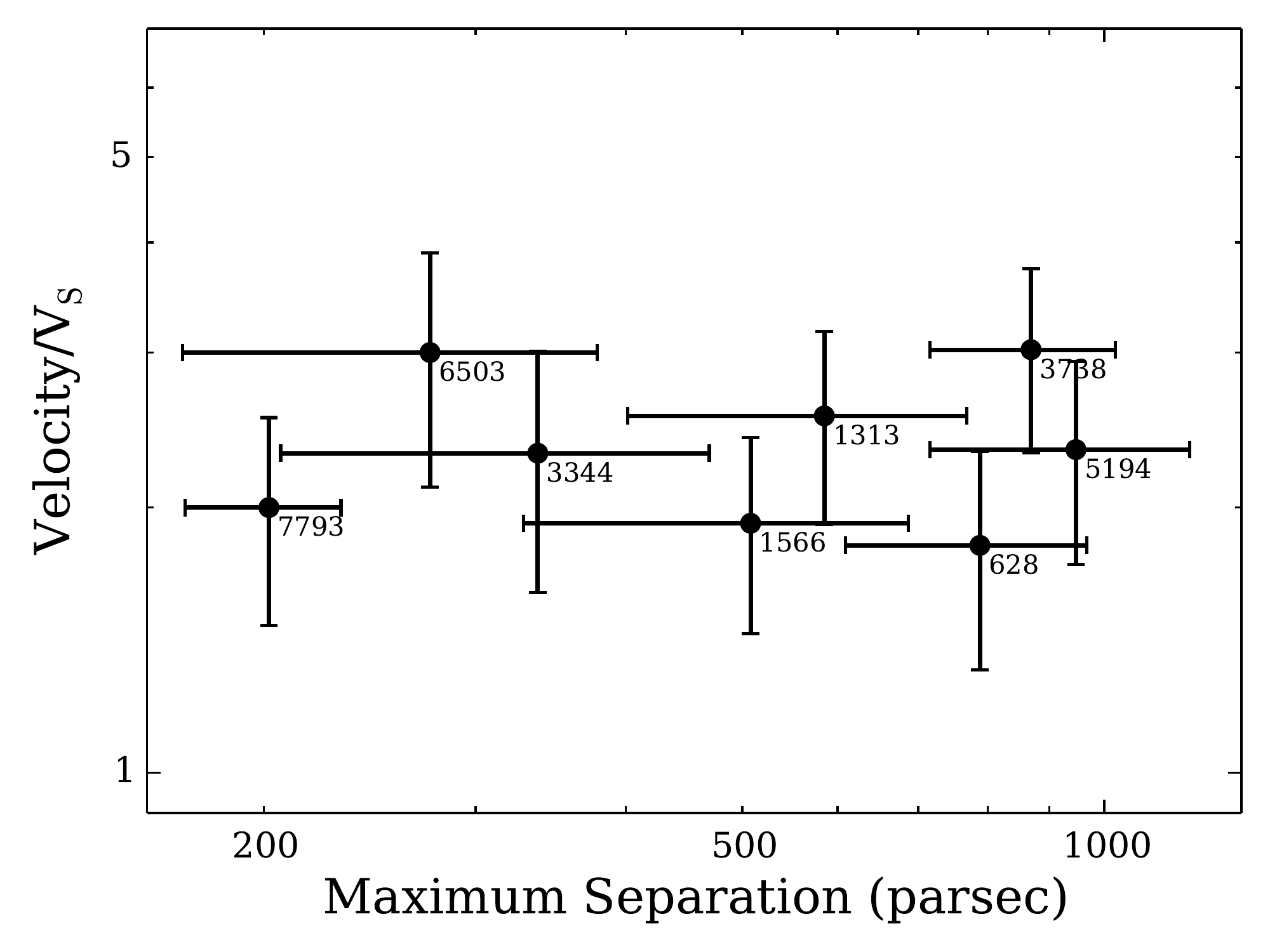}
\caption{
The velocity at $R_{\rm max}$ divided by the velocity difference due to shear $V_{\rm S}$ as a function of the turnover size between cluster pairs $R_{\rm max}$.  The ratio $Velocity/V_{\rm S}$ is independent of $R_{\rm max}$, implying that shear is responsible for determining the maximum size of star-forming regions.  Error bars display the standard 1$\sigma$ error in the mean.
\label{fig:shear}}
\end{figure}

The velocity from the $\Delta t - R$ relation, when normalized to the velocity component produced by shear (Figure \ref{fig:shear}), is independent of the turnover size $R_{\rm max}$.  Thus the maximum speed associated with the largest structure of star formation in a galaxy is linked directly to the velocity difference from shear within the same structure.  These results indicate that while turbulence is quite possibly the dominant process defining the $\Delta t - R$ relation, there are dependencies on the environment of the host galaxy, which affect the measured maximum sizes, age-differences, and velocities. 

\begin{figure*}
\includegraphics[scale=0.48]{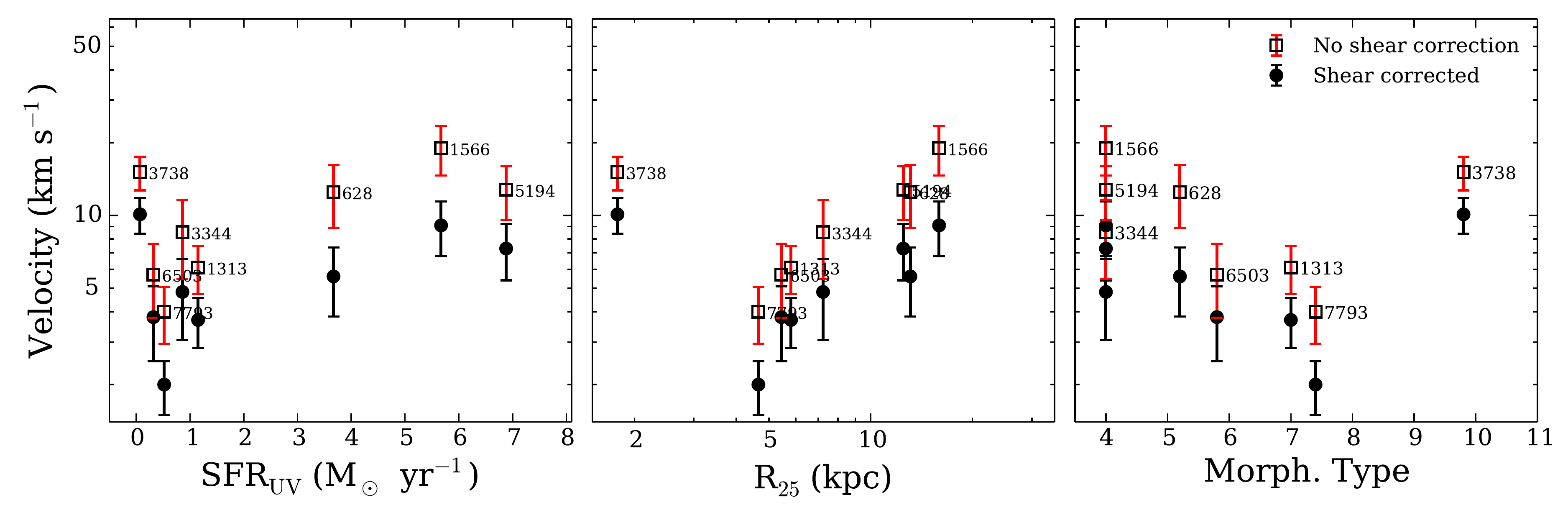}
\caption{
The velocity at $R_{\rm max}$ versus the UV SFR (left), R$_{25}$ optical radius (center), and morphology T--type (right).  Open squares with red error bars have not been corrected for shear while black solid circles with error bars have been corrected for shear.  Larger spiral galaxies and higher SFRs correlate with larger velocities; the Irregular galaxy NGC 3738 does not follow the spirals' trend.  
\label{fig:absmag}}
\end{figure*}

\begin{figure}
\includegraphics[scale=0.42]{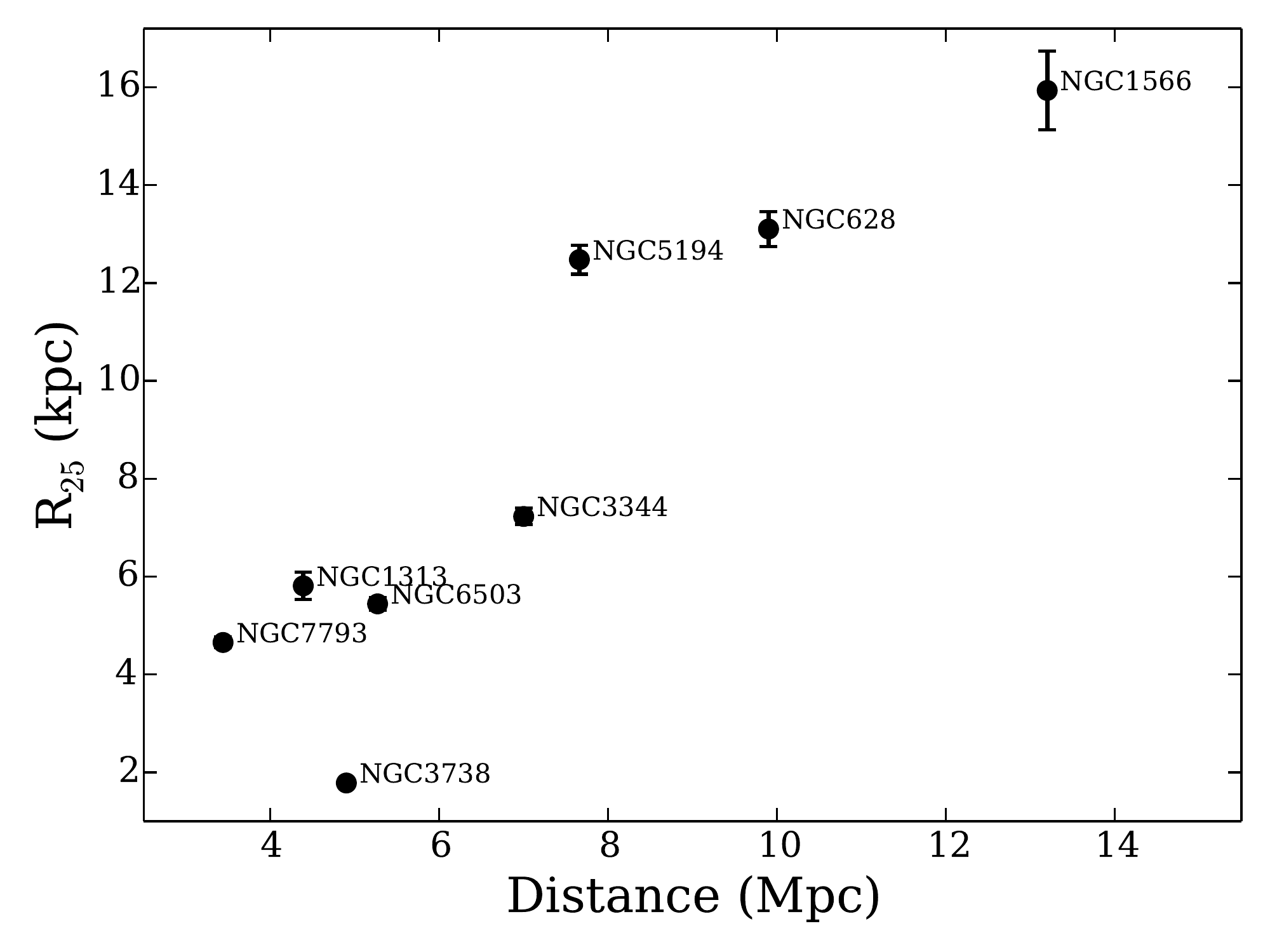}
\caption{
The R$_{25}$ radius as a function of the distance for each galaxy.  The largest galaxies, and therefore, the largest traveling velocities calculated from the $\Delta t - R$ relation, are found at the largest distance within our survey.  
\label{fig:bias}}
\end{figure}

In Figure \ref{fig:absmag}, we show trends of the velocity at $R_{\rm max}$ for each galaxy with the dust-corrected star formation rate (SFR), R$_{25}$ optical radius, and morphological Hubble T--type.  We find that larger spirals (correlation coefficient $r=0.98$) with higher SFRs ($r=0.86$) and smaller T--type morphologies ($r=0.83$) exhibit larger $R_{\rm max}$ velocities, although the trends with SFR and T-type are weak.  This is not unexpected as larger galaxies are known to have larger star-forming complexes \citep{elmegreenetal96}.  The irregular galaxy, NGC 3738, does not follow the trend that we observe for spirals.  Because the cluster catalogs are complete to clusters more extended than $R_{eff} \sim 1$~parsec \citep{adamo17} at a distance of 10 Mpc, and we expect this to apply to all galaxies excluding NGC 1566 (13 Mpc), we do not expect any significant bias in our cluster analysis due to the distance of the galaxies.  The apparent correlation between an increase in the velocity (or maximum size $R_{\rm max}$) with increasing galaxy distance (Table \ref{tab2}) is an observational bias in our sample as the smallest galaxies in our sample are at the closest distances (Figure \ref{fig:bias}).

\subsection{Cluster Evolution}\label{sec:dissolution}
Clusters are disrupted rapidly due to evaporation, merging, and tidal field interactions \citep[for a review, see][]{portegieszwart10}, removing up to 90\% of clusters within each age dex.  At older ages, cluster samples start to suffer from incompleteness due to evolutionary fading and rising mass completeness limits at ages older than a few hundred Myr \citep[e.g., ][]{fouesneau14,adamo17}.  Incompleteness due to embedded clusters, confusion with associations, and infant mortality is important in the age range 1--10 Myr \citep{gieles11}.  We can test the impact of cluster evolution on our results by inspecting the $\Delta t - R$ relation for clusters with ages above $\sim$10 Myr and for ages up to 100 Myr compared to 300 Myr.  We compute the $\Delta t - R$ relation in this manner for four of our galaxies with the most populous cluster catalogs in our sample with the best number statistics available: NGC 1313, NGC 5194, NGC 628, and NGC 1566.

\begin{figure*}
\includegraphics[scale=0.43]{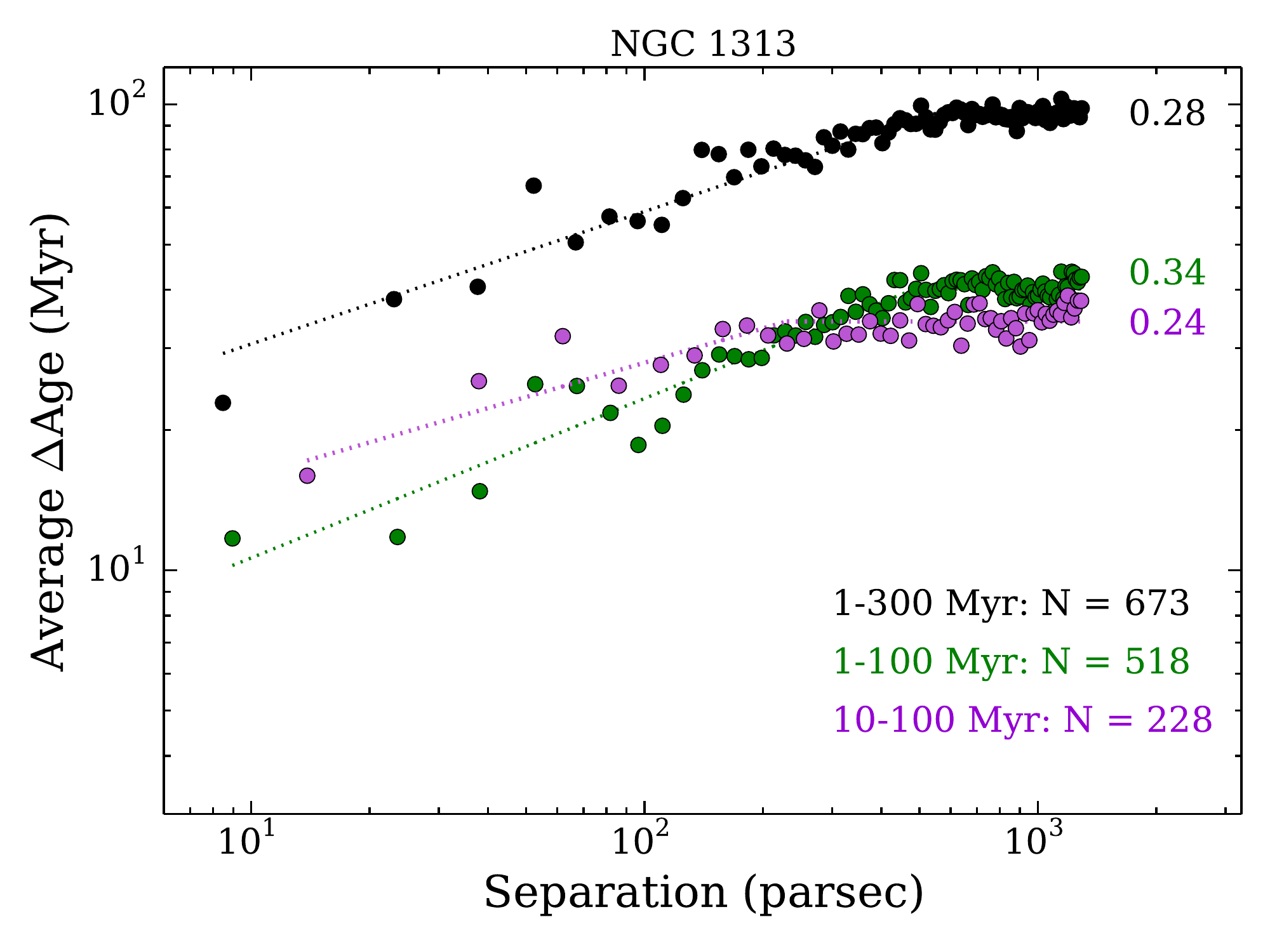}
\includegraphics[scale=0.43]{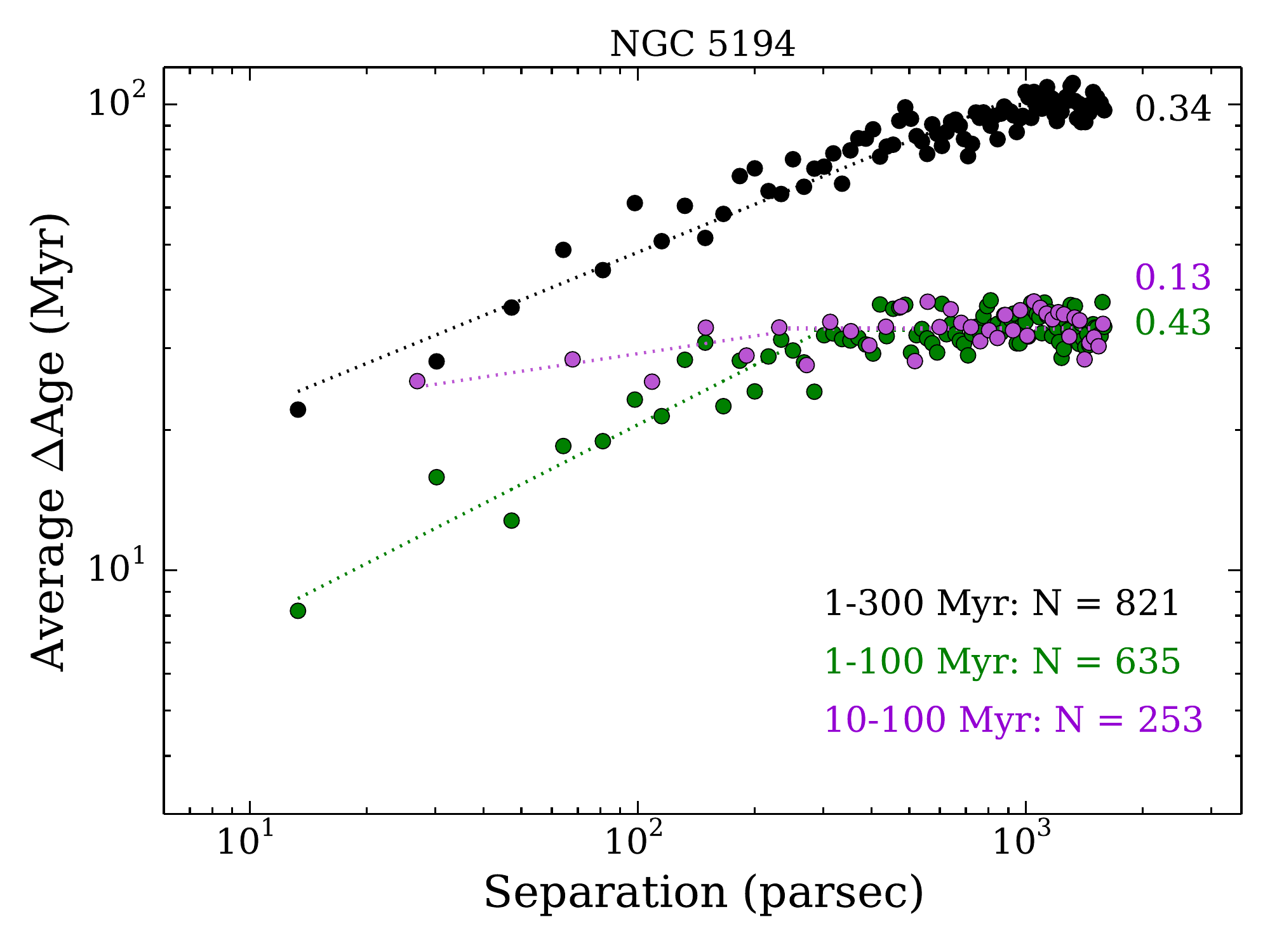} \\
\includegraphics[scale=0.43]{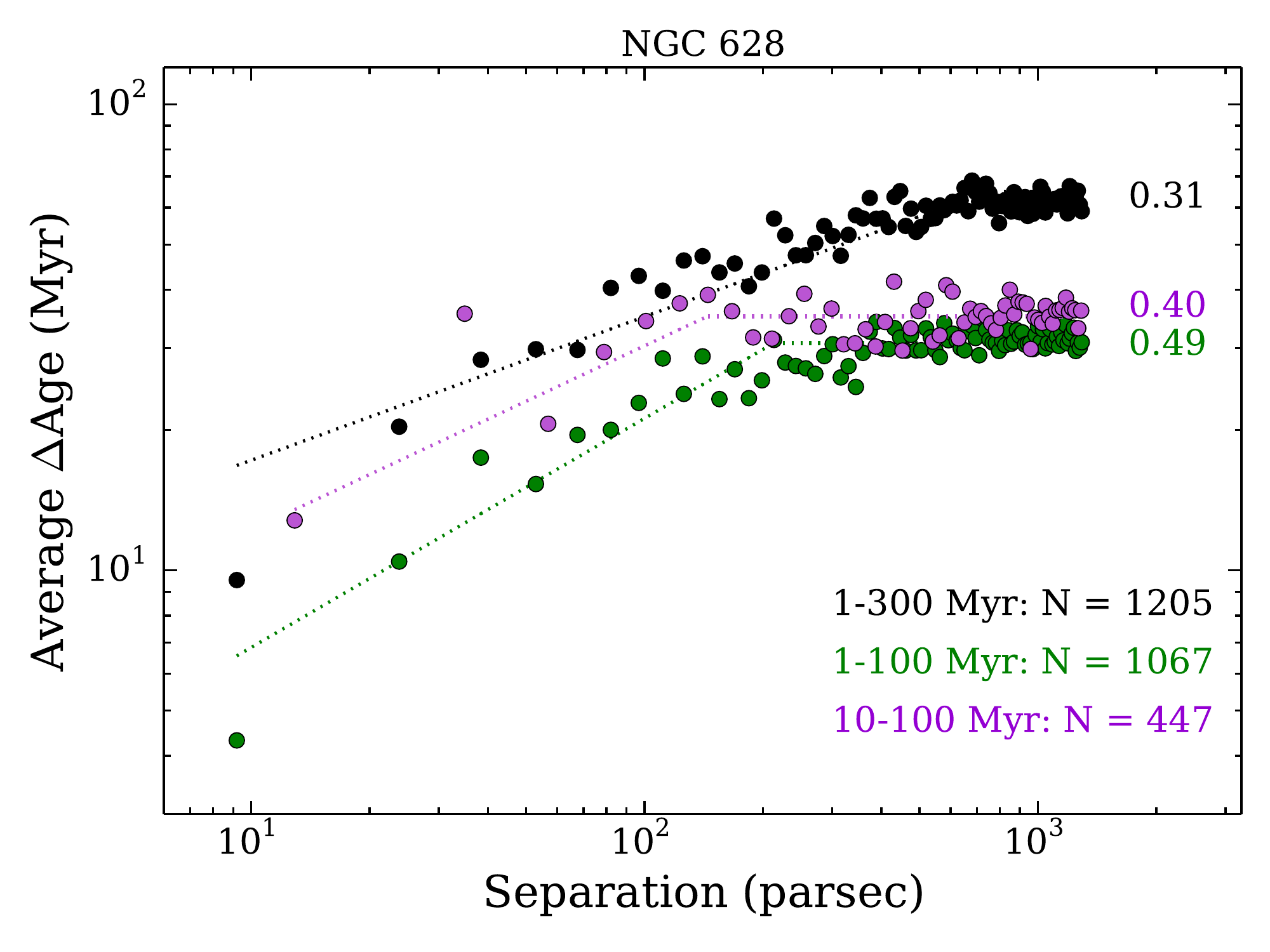}
\includegraphics[scale=0.43]{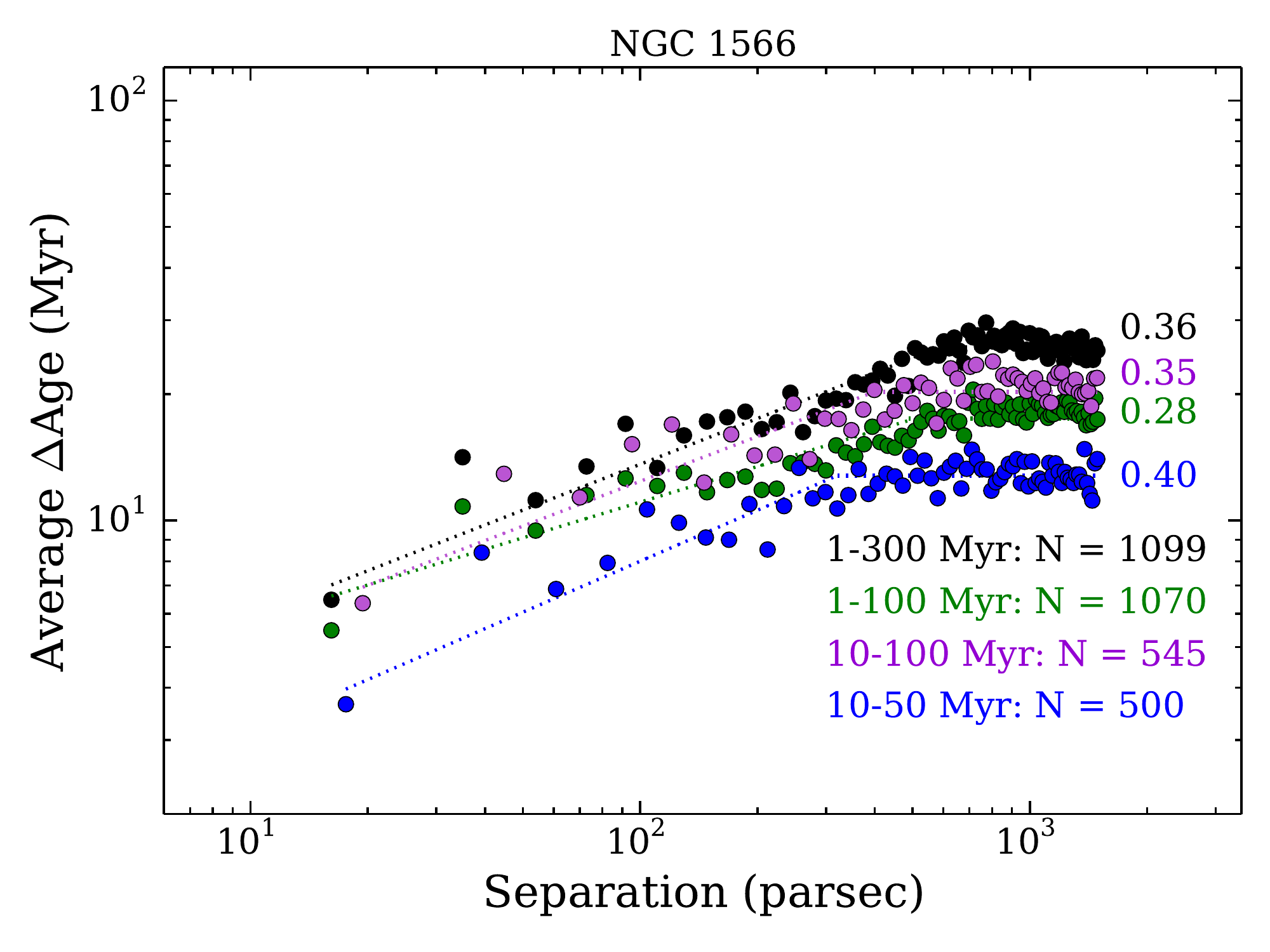}
\caption{
The age difference between cluster pairs as a function of separation in logarithmic scale between the cluster pairs of four galaxies, divided into three different age ranges to test for the effect cluster evolution on the $\Delta t - R$ relation for an aging cluster population: 1--300 Myr (black), 1--100 Myr (green), and 10-100 Myr (purple).  NGC 1566 contains enough clusters to also investigate the trend at ages 10--50 Myr (blue).  The range 10-100 Myr is selected to avoid the effects of violent gas disruption (ages $<$10 Myr) and cluster dissolution and fading (ages $\gtrsim 100$ Myr).  A fit to the $\Delta t - R$ relation is performed and the numbers to the right of each age range show the value of the slope $\alpha$, with the full power-law fits listed in Table \ref{tab3}.  The scatter present in the 10-100 Myr range is significant and should be represented with a single power-law; however we perform the fits in this age using a double power law like in the other cases, to enable direct comparisons among the three age ranges considered.  Older clusters result in larger, more extended star-forming sizes, increasing the size of $R_{\rm max}$.  Cluster dissolution and fading also has the greatest impact for clusters older than a couple hundred Myr, increasing the average separation between pairs and flattening the slope in the $\Delta t - R$ correlation as clusters age.  
\label{fig:agedependency}}
\end{figure*}

Based on the considerations above, we expect that clusters with ages of 10--100 Myr are those for which the effect of both cluster disruption, fading, and dynamical mass loss will have minimal impact on the $\Delta t - R$ relation.  We show this in Figure \ref{fig:agedependency} along with the $\Delta t - R$ relation for the age range 1--300 Myr and 1--100 Myr.  Table \ref{tab3} lists the power-law fits of the four galaxies available for all the different age ranges using logarithmic bins like in Figure \ref{fig:agespread}.  The best-fit values (e.g., $R_{\rm max}$) do change between the two binning methods of Figure \ref{fig:agespread} and Figure \ref{fig:agedependency}, but the values are well within the error.  The scatter is significant in the age range of 10--100 Myr because of small number statistics, and the correlation should be represented with a single power-law.  However, we elect to keep the double power-law fit in order to enable easy comparison of this age bin with the other two.

Evolutionary effects on the clusters already appear to have impacted the age--separation results by 300 Myr, flattening the slopes and increasing the characteristic star formation timescale compared to clusters that are only 100 Myr old in NGC 628, NGC 5194, and NGC 1313.  On the other hand, NGC 1566 shows a different trend where the measured slope decreases when we lower the age limit from 300 to 100 Myr.  However, the slope of NGC 1566 marginally increases when we further decrease the age limit to the range 10--50 Myr, the only system were we have enough statistics to test the relation in the range for clusters in this smaller age range.  The ability to perform tests that pinpoint the age where evolutionary effects become important depend on the size of the available catalog.  Not accounting for the effect of cluster evolution may act to decrease the underlying slope.  

An inspection of the clusters in the range 10-100 Myr compared to 1--100 Myr in Figure \ref{fig:agedependency} -- effectively removing clusters that are subject to violent gas expulsion \citep[e.g.,][]{baumgardt13} -- show that the $\Delta t - R$ relation becomes shallower as clusters younger than 10 Myr are removed.  The relation steepens marginally within NGC 1566, but as the total age range is narrow, the scatter is considerable and we do not consider it a significant effect.  

The clusters in the range 10--100 Myr for both NGC 1313 and NGC 5194 show an extremely shallow relation compared to what is observed for the rest of the systems.  We are limited by small number statistics for these two systems for this age range, having only 228 and 253 clusters, respectively, compared to 447 and 545 clusters for NGC 628 and NGC 1566, respectively.  Thus, the scatter is significant and part of the observed shallowing of the slope is likely due to small number statistics for the shortest spatial scales with the youngest age differences. 

In general, the slope steepens when we reduce the upper age limit.  \citet{delafuentemarcos09} also found that the MW cluster pairs exhibit a steepening slope after correcting for cluster dissolution and incompleteness.  Our catalogs are most sensitive to clusters with young ages: between 45\%--66\% of the star clusters in each galaxy have ages $\lesssim$10 Myr.  The effect of removing clusters with ages $<$10 Myr impacts the age--separation relation in each galaxy differently, but in all cases, it is apparent that the youngest, most recent star forming regions drive the observed $\Delta t - R$ relation at the smallest scales.  While the initial star cluster formation may be driven by turbulence, after the star clusters have aged just a few tens of Myrs, the imprint of turbulence on their age--separation relation appears to have diminished.  

The recovered slope of NGC 628 exhibits a significant flattening for clusters older than 100 Myr which is not observed for the other galaxies.  This agrees with the findings of \citet{adamo17} that cluster disruption is not important for ages between 10 and 200 Myr in NGC 628, hence the dramatic change in the slope from 100 to 300 Myr.  Both NGC 5194 \citep{gieles05} and NGC 1566 \citep{hollyhead16} also show slow dissolution rates of $\sim 100$~Myr for their cluster populations, unlike NGC 1313, which exhibits a fast cluster disruption timescale of 25~Myr \citep{pellerin07}.   

\begin{figure}
\includegraphics[scale=0.42]{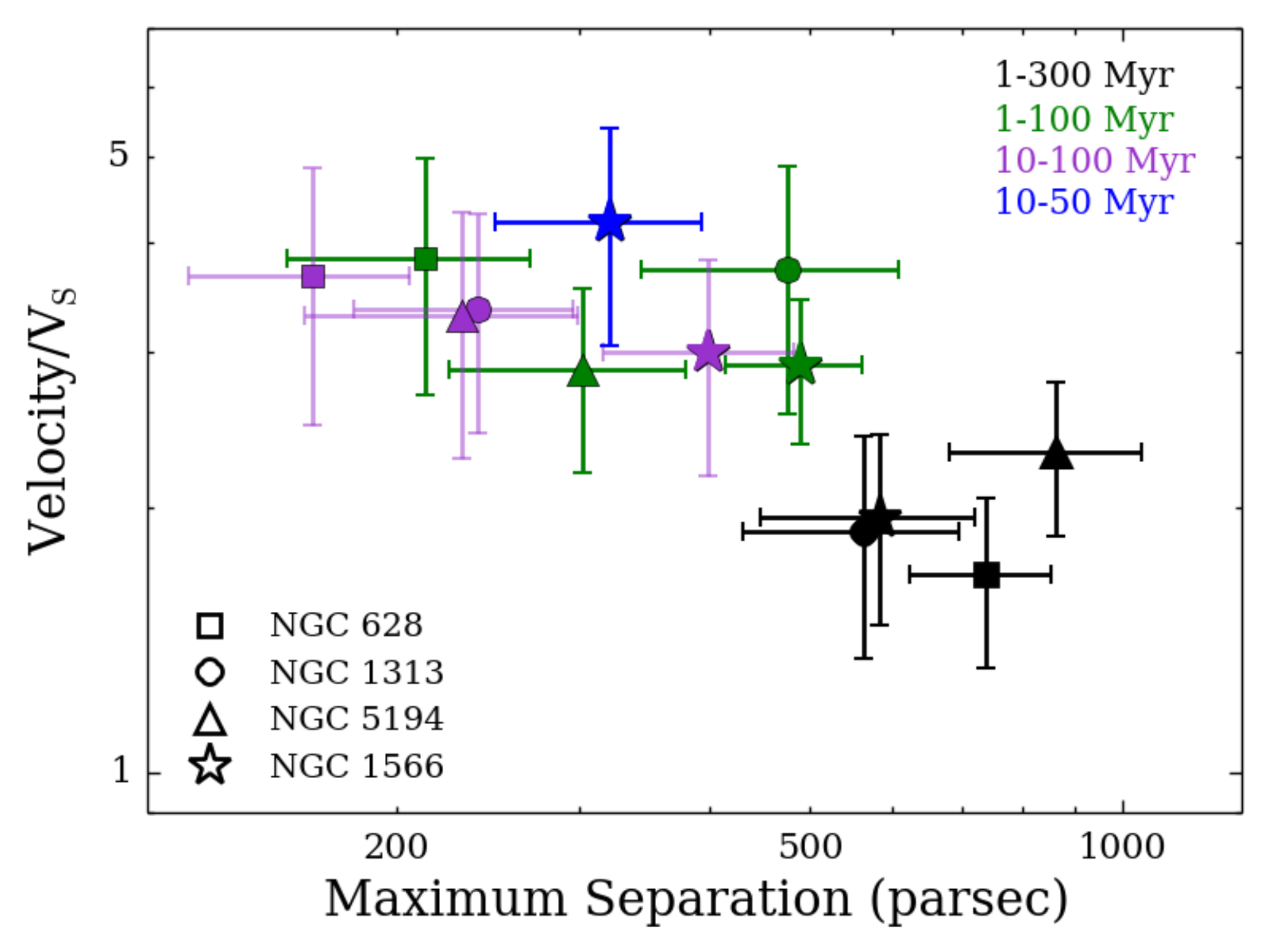}
\caption{
Similar to Figure \ref{fig:shear}, the velocity at $R_{\rm max}$ divided by the velocity difference due to shear $V_{\rm S}$ as a function of the turnover size between cluster pairs $R_{\rm max}$ calculated from the different age ranges from Figure \ref{fig:agedependency}.  The black symbols show the $V/V_{\rm S}$ ratio for clusters in the age range 1--300 Myr, green show the range range 1--100 Myr, purple show the age range 10--100 Myr, and blue show the age range 10--50 Myr (only for NGC 1566).  Individual galaxies are represented by different symbols:  NGC 628 (squares), NGC 1313 (circles), NGC 5194 (triangles), and NGC 1566 (stars).  The range 10--100 Myr would be better represented by a single power-law (Figure \ref{fig:agedependency}); however, we elect to fit that age range with a double power-law to enable direct comparison with the other two age bins.  We highlight this choice for the 10--100 Myr age bins by marking the error bars for this age bin with lighter lines.  For a given galaxy, younger (i.e., smaller) star-forming regions have a total traveling velocity which is a few times larger than the shear, while approaching the V$_{\rm S}$ value as the maximum region size increases.  Within each age range, however, V/V$_{\rm S}$ remains constant from galaxy to galaxy.  Error bars display the standard 1$\sigma$ error in the mean.
\label{fig:shear_multi}}
\end{figure}

Figure \ref{fig:shear_multi} shows the ratio of the traveling velocity to the velocity difference due to shear $V_{\rm S}$ as a function of the turnover size for the different age ranges of Figure \ref{fig:agedependency}.  The four galaxies with enough clusters to perform an analysis of the dependence of the $\Delta t - R$ relation with age are also the largest galaxies with the largest derived $R_{\rm max}$ value.  While there is a general trend for a higher ratio of $V/V_{\rm S}$ at the turnover point when tracing younger star-forming regions within a given galaxy (i.e., a lower contribution of shear within the youngest, smallest complexes), selecting the same age range for all galaxies results in a flat trend between $V/V_{\rm S}$ and $R_{\rm max}$.  As long as we compare the results using the same age range for the clusters, the ratio $V/V_{\rm S}$ appears to be constant across the selected cluster age range.  

The difference in the observed slopes or maximum age is not an artifact of galaxy distance because, comparing Tables \ref{tab1} and \ref{tab2}, there is no correlation between maximum correlation age and distance.  Shear inside the galaxy may be an important factor in determining the duration of star-formation and size of regions, as shown in Section \ref{sec:shear} and Figures \ref{fig:shear} and \ref{fig:shear_multi}.  

\begin{deluxetable}{lccccc}
\tablecolumns{6}
\tabletypesize{\footnotesize}
\tablecaption{Age Difference and Spatial Separation Results for Varying Age Ranges\label{tab3}}
\tablewidth{0pt}
\tablehead{
\colhead{Galaxy} 	& 
\colhead{$A_1$} & 
\colhead{Slope} 		&
\colhead{Max Age} &
\colhead{$R_{\rm max}$}&
\colhead{V/V$_{\rm S}$} 
\\
\colhead{} 					& 
\colhead{(Myr)} 		& 
\colhead{($\alpha$)} 	&
\colhead{(Myr)} 			& 
\colhead{(pc)} 			&
\colhead{} 
\\
\colhead{(1)} & 
\colhead{(2)} & 
\colhead{(3)} & 
\colhead{(4)} & 
\colhead{(5)} & 
\colhead{(6)} 
}    
\startdata    
\multicolumn{6}{c}{10--50 Myr}  \\
\cline{1-6} \\     
N1313 & \multicolumn{5}{c}{\nodata} \\
N5194 & \multicolumn{5}{c}{\nodata} \\
N628   & \multicolumn{5}{c}{\nodata} \\
N1566 & 1.3(0.3)	& 0.40(0.09)	& 13(3)  & 326(72)  & 4.2(1.2) \\
\cline{1-6} \\ 
\multicolumn{6}{c}{10--100 Myr$^{\rm a}$ \footnotetext{ The clusters in the 10--100 Myr range of Figure \ref{fig:agedependency} are best-described with a nearly flat, single power-law with slope with no breakpoint.  However, in order to compare the ratio of the velocity from shear in Figure \ref{fig:shear_multi}, we keep the double power-law fit.}  }  \\
\cline{1-6} \\  
N1313 & 9(2)	 		& 0.24(0.06)	& 34(8) 	& 239(57) 	& 3.4(0.9)\\
N5194 & 16(4)  		& 0.13(0.03)	& 33(9) 	& 258(68)	& 3.3(1.0)\\
N628   & 5.0(1.3) 	& 0.40(0.11) 	& 35(10)	& 141(40)	& 3.7(1.2)\\
N1566 & 2.4(0.5) 	& 0.35(0.07) 	& 20(4) 	& 408(93)	& 3.0(0.8)\\
\cline{1-6} \\ 
\multicolumn{6}{c}{1--100 Myr}  \\
\cline{1-6} \\ 
N1313 & 4.8(1.3) & 0.34(0.07) &40(11)& 486(132)	& 	3.7 (1.2) \\
N5194 & 2.8(0.7) & 0.43(0.08) & 33(6) & 294(77) 	&	2.9(0.7) \\
N628   & 2.2(0.6) & 0.49(0.13) & 31(8) & 211(56) 	&	3.8(1.1) \\
N1566 & 3.0(0.4) & 0.28(0.04) & 18(2) & 493(73) 	& 	2.9(0.5) \\
\cline{1-6} \\ 
\multicolumn{6}{c}{1--300 Myr}  \\
\cline{1-6} \\ 
N1313 & 16(3)			& 0.28(0.07)	& 96(23)	& 560(132)	& 1.9(0.5)\\
N5194 & 10.0(1.3)		& 0.34(4)			& 99(13)	& 842(180) & 	2.3(0.5)\\
N628   & 7.8(1.2)		& 0.31(0.05)	& 64(10)	& 736(114)	& 1.7(0.4)\\
N1566 & 2.1(0.4)		& 0.40(0.07)	& 26(5) 	& 590(136) & 1.9(0.5)
\enddata
\tablecomments{
Power-law fits for the logarithmic separation bins for the galaxies in Figure \ref{fig:agedependency}. 
Columns list the: 
(1) galaxy name, ordered by increasing distance; 
(2) intercept $A_1$; 
(3) slope $\alpha$; 
(4) $A_2$, the average age difference between cluster pairs at the turnover, $R_{\rm max}$; 
(5) size $R_{\rm max}$ of the star forming region at the turnover for the age range from Eq. \ref{eq:1}; and 
(6) the ratio of the traveling velocity to the velocity different from shear (Figure \ref{fig:shear_multi}).  Each value is listed for four different age ranges:  10--50 Myr, 10--100 Myr, 1--100 Myr, and 1--300 Myr.  Only NGC 1566 was investigated for the range 10--50 Myr.
Numbers in parentheses indicate $1\sigma$ uncertainties in the final digit(s) of listed quantities, when available. 
}
\end{deluxetable}

\subsection{Binning Method and the Dependence of $R_{\rm max}$ on Age}\label{sec:agedependency}
In this section, we investigate the choice of bin size and method of binning on the $\Delta t - R$ results, specifically the sensitivity of the location of the turnover point $R_{\rm max}$ to the binning selection.

The total spread of possible age differences between cluster pairs is significant, especially at the largest separations.  While the upper envelope of age differences only increases with increasing separation, there will always be cluster pairs with $\Delta t = 0$ values at all separations.  As performed in both \citet{efremov98} and \citet{delafuentemarcos09}, we bin the data, finding that the best representation of the underlying distribution is when the binning method is performed in equal logarithmic-spacing, as shown in Figure \ref{fig:agespread}.  However, in order to fully understand how our choice of binning methods may potentially impact the location of $R_{\rm max}$, which drives the resulting velocity at that location, we also compute the $\Delta t - R$ correlation for regular intervals of separations of 20 parsecs for our selected age range of 1--300 Myr.  

Figure \ref{fig:sto} shows the results: as expected, when we increase the number of bins, the scatter between individual bins increases.  Additionally, choosing the bins such that they are for regular intervals of separation greatly limits the resolution at small separations due to sparse numbers of clusters.  The derived values of $R_{\rm max}$ remain consistent relative to those listed in Table \ref{tab2}.  The $\Delta t - R$ relations in Figure \ref{fig:agedependency} for different age ranges can be used to investigate the dependency of the size of a star-forming region, $R_{\rm max}$.  

For all galaxies, when the maximum age of the clusters is lowered and we only consider clusters younger than 100 Myr for the $\Delta t - R$ relation, we find that the size of the star-forming region $R_{\rm max}$ decreases (Figure \ref{fig:agedependency} and Table \ref{tab3}).  This is expected as the youngest clusters will trace the regions of the most recent star formation, occurring at the smallest size scales.  Thus, increasing the age of the star-forming region increases the extent of the structure that the star clusters are tracing, and as expected, increases the maximum age of the region as well.  Only half our sample have large enough cluster catalogs to investigate the age-dependency on the $\Delta t - R$ relation.  The changing of the $\Delta t - R$ relation with different age ranges provides important insight on the effect of cluster dissolution and fading on the results, as discussed in the previous section.

\subsection{Sensitivity of the $\Delta t - R$ Relation on Stochastically Derived Properties}\label{sec:sto}
As mentioned in Section \ref{sec:completeness}, the cluster properties for two of the LEGUS galaxies, NGC 7793 and NGC 628, have been derived through the implementation of SLUG \citep{krumholz15}.  SLUG provides stochastically derived posterior probability distribution functions (PDFs) for ages, masses, and extinctions of the clusters, assuming different priors for both the cluster mass function and dissolution rate.  We utilize the full posterior PDFs of the cluster candidates in NGC 7793 and NGC 628 to investigate the implications on the $\Delta t - R$ relation of deriving the ages stochastically compared to traditionally deterministic SED modeling.  We ensure that the underlying physical assumptions (e.g., metallicity, dust attenuation models) are consistent between the two codes.
 
\begin{figure*}
\includegraphics[scale=0.44]{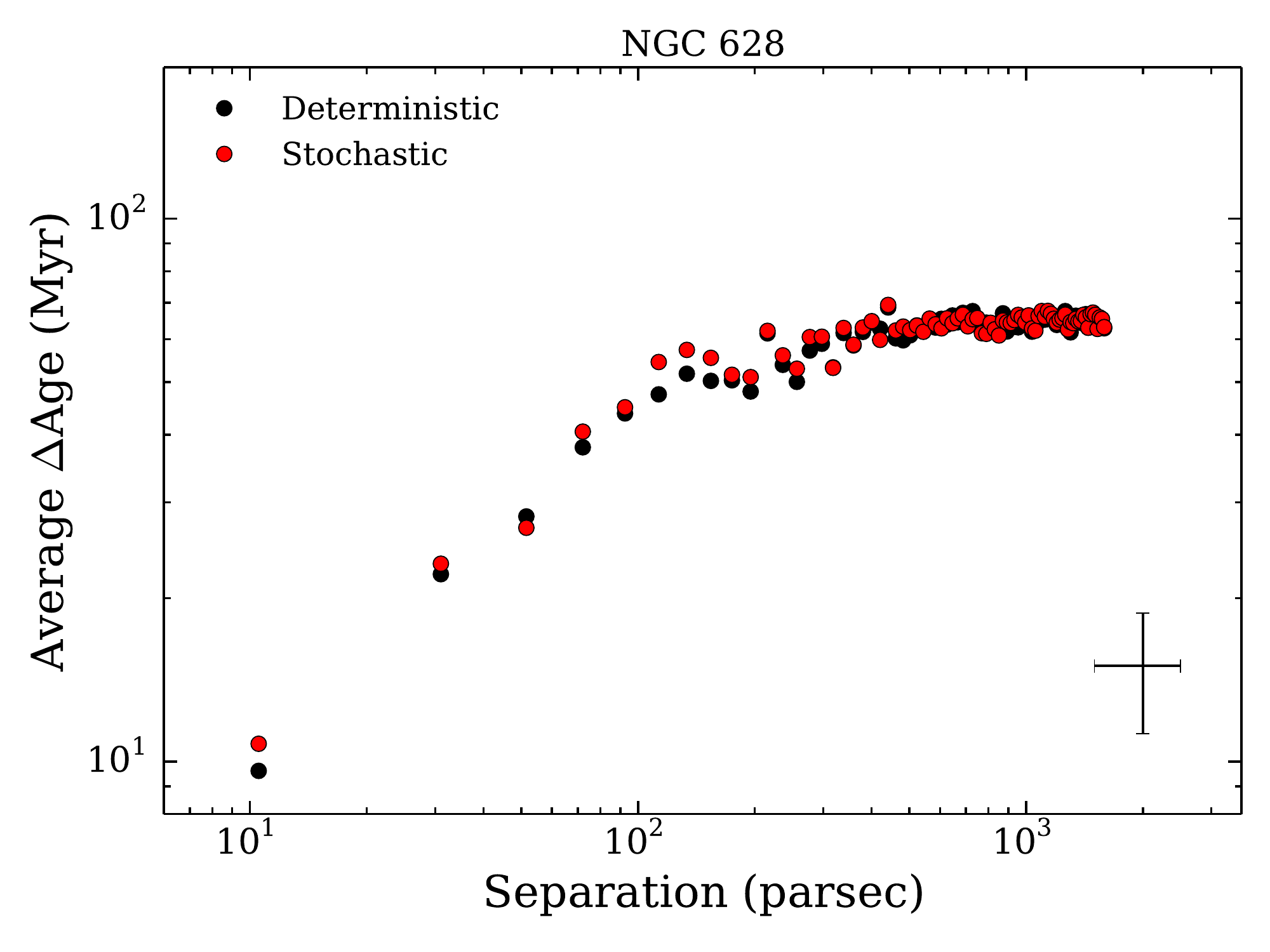}
\includegraphics[scale=0.44]{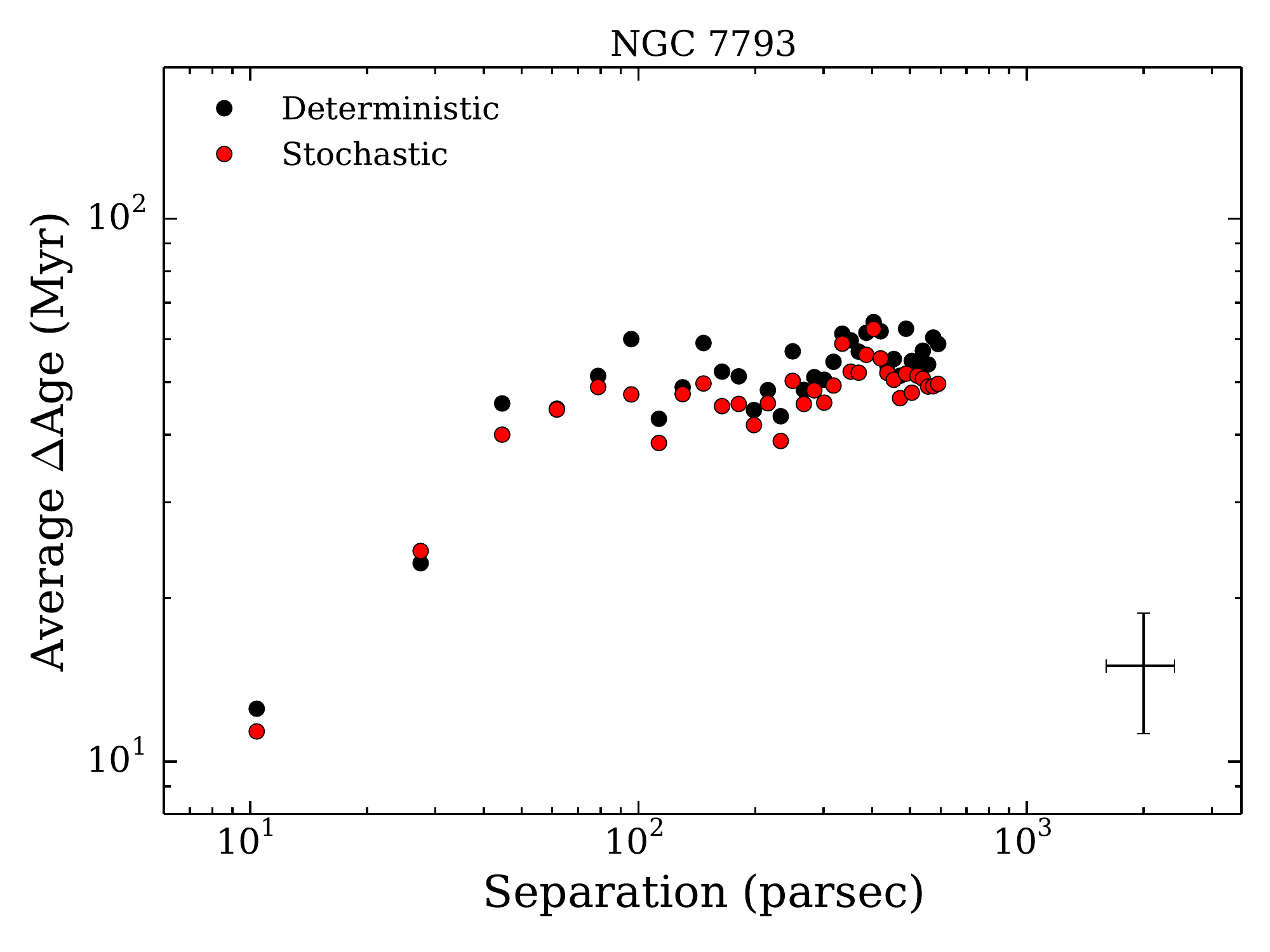}
\caption{
The age difference between cluster pairs as a function of separation of the cluster-rich galaxy NGC 628 (left) and the cluster-poor galaxy, NGC 7793 (right).  We show the average age difference $\Delta t$ as a function of their physical separation $R$ for linear intervals of separation of 20 parsecs.  
The black points show the results if the ages are determined with deterministic models and the red points show the results for the ages derived with stochastic modeling (Section \ref{sec:sto}).  The bottom right shows the average error for each point.  The difference between the $\Delta t - R$ relation for the stochastic models are well within the scatter of the deterministic models.  
\label{fig:sto}}
\end{figure*}

Figure \ref{fig:sto} shows the $\Delta t - R$ relation for NGC 7793 and NGC 628 where the ages are determined both from deterministic techniques and with the stochastic modeling code SLUG \citep{krumholz15}.  We take the stochastic age of each cluster as the peak of the marginal posterior probabilities.  


We find that the global trends from ages determined stochastically are similar to what is observed for cluster properties derived from traditional SED fitting techniques.  There is a slight decrease in the average age separations for clusters with properties derived stochastically within NGC 7793 and no difference is observed within NGC 628.  Any difference is well within the uncertainties on the age differences as determined from deterministic models.  We conclude that there is no concern of any bias introduced in our results that is caused by the traditional deterministic fitting procedures.  


\subsection{Randomization Tests}\label{sec:tests}
\subsubsection{Shuffling the Ages}\label{sec:ageshuffle}
\begin{figure*}
\includegraphics[scale=0.44]{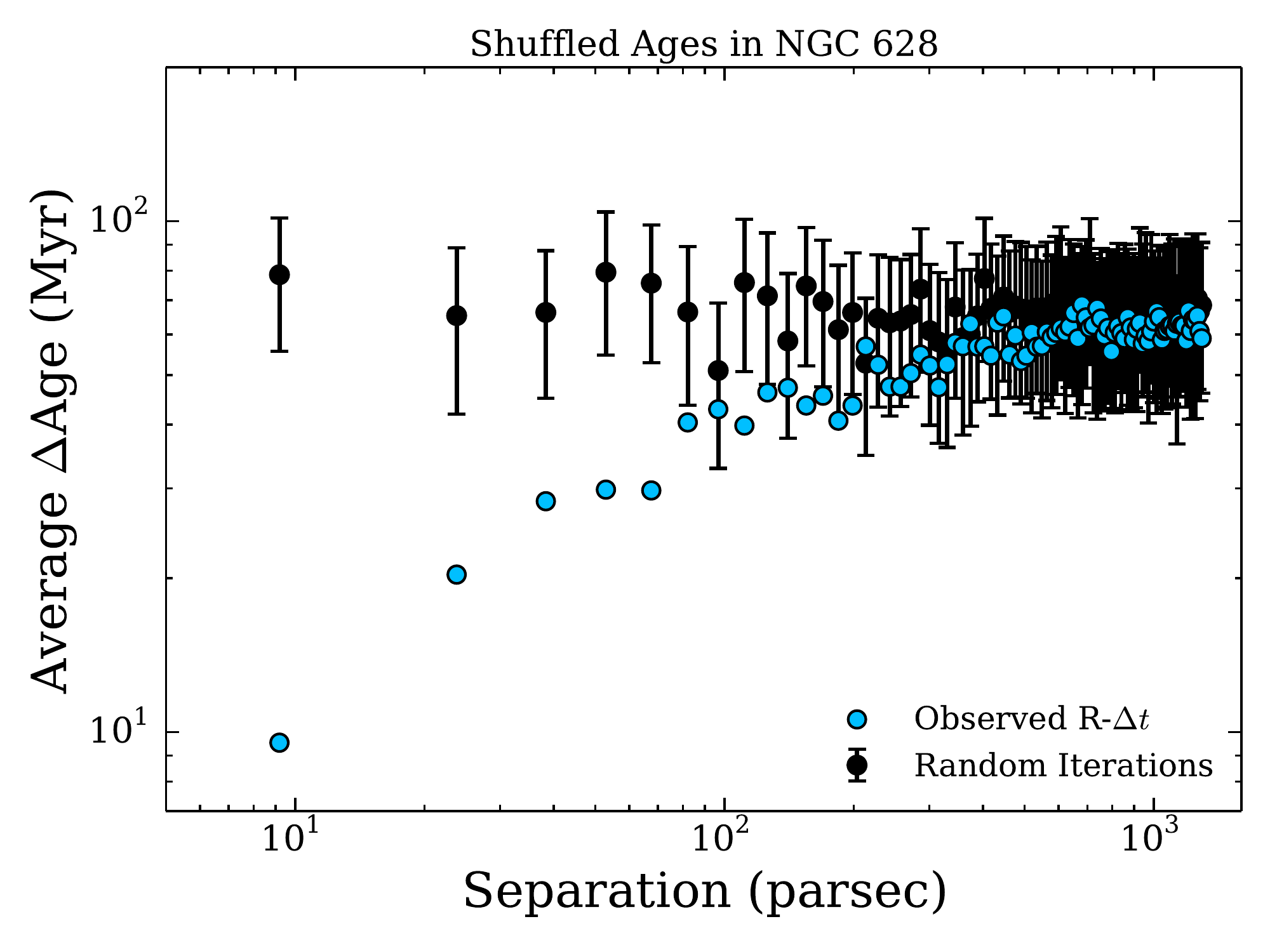}
\includegraphics[scale=0.44]{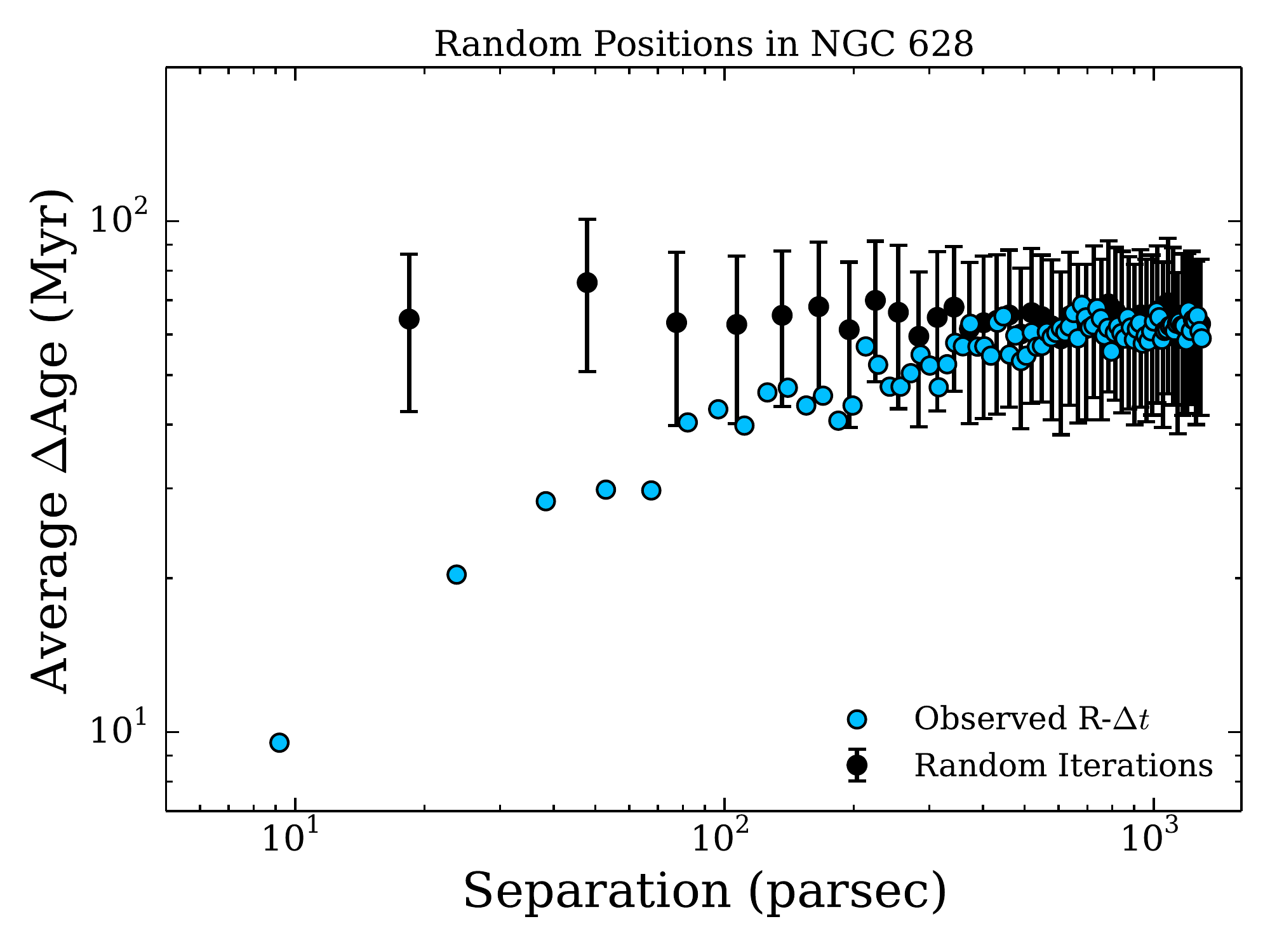}\\
\includegraphics[scale=0.44]{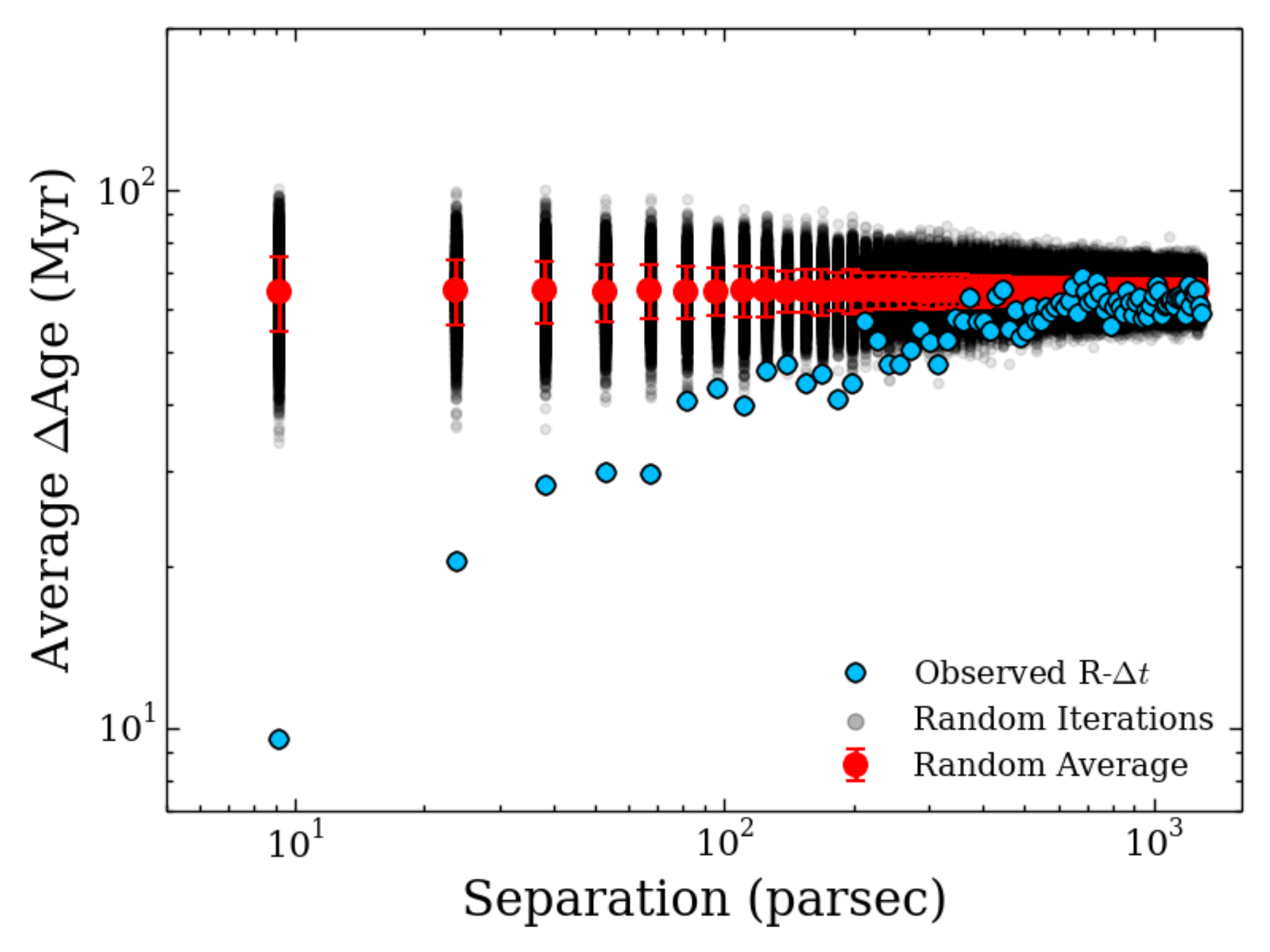}
\includegraphics[scale=0.44]{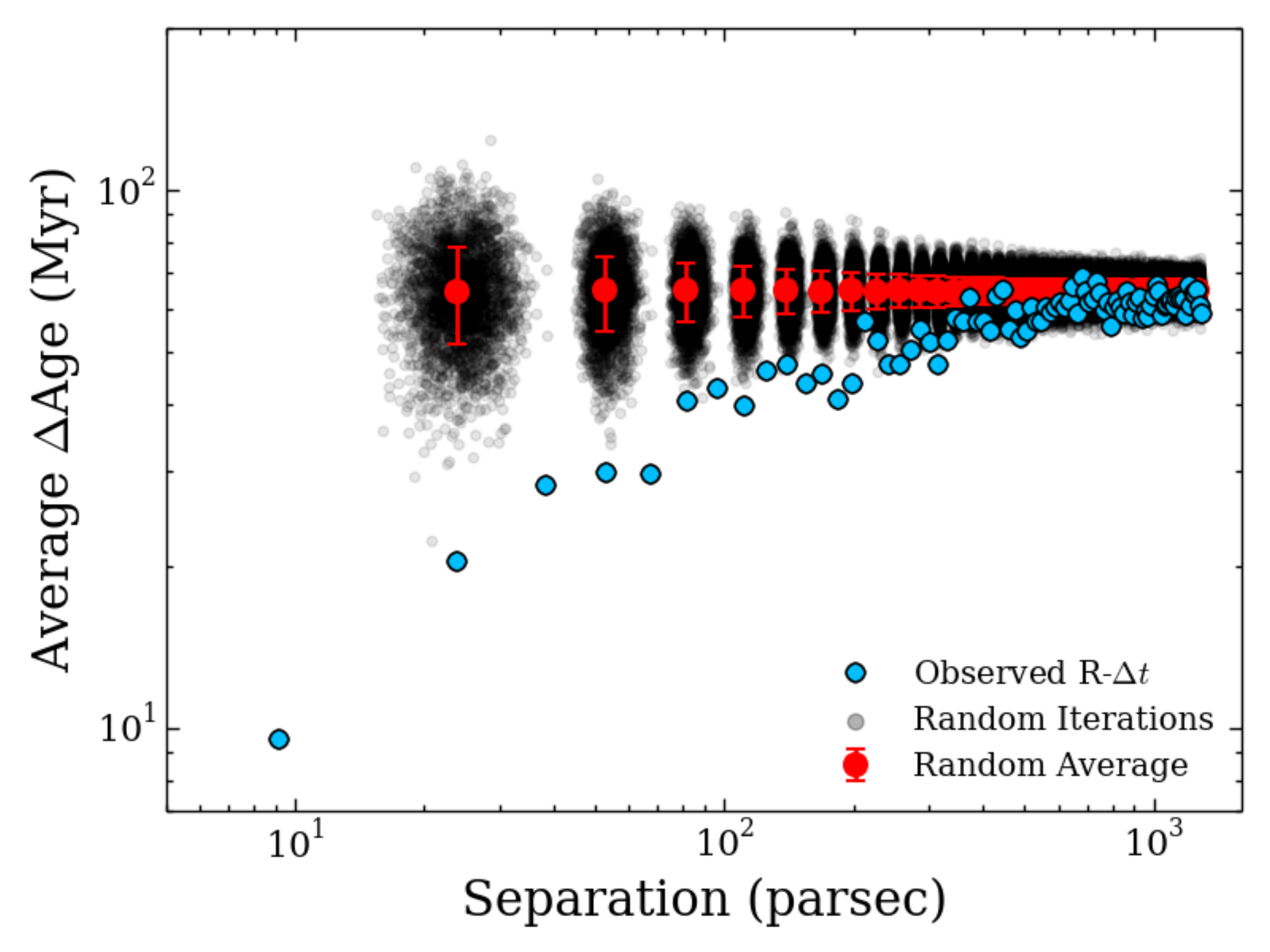}
\caption{
The age difference between cluster pairs as a function of separation in logarithmic scale between the cluster pairs of NGC 628.  Blue points show the trend for the real data of NGC 628.  {\it Top left}:  One example of the $\Delta t - R$ relation where we randomly shuffle the ages among the real cluster positions.  {\it Top Right}:  One example with randomizing the cluster positions within NGC 628 (see Section \ref{sec:randompos}).  {\it Bottom Left}: The gray points show 50,000 iterations where cluster ages are shuffled randomly among the real cluster positions, and the red points show the average of the shuffled points with the 1$\sigma$ error in the mean.  {\it Bottom Right}: The gray points show 15,000 iterations when cluster positions are randomized within the galaxy, and the red points show the average of the randomized points with the 1$\sigma$ error in the mean.  There are a lack of close cluster pairs with randomized positions as real cluster positions are highly clustered, especially at young ages.  The correlation disappears for randomized ages and positions across all pair separation lengths, suggesting that the $\Delta t - R$ trend found in Figure \ref{fig:agespread} and shown in blue for the clusters is statistically significant and determined by a real physical effect.  
\label{fig:randomizeage}}
\end{figure*}

We perform two tests to assess whether the correlation found between the age and separation of cluster pairs is truly statistically significant or is an effect of our sample's luminosity-limited selection.  In the first test, using the 1205 star clusters in NGC 628, we keep the real cluster positions and shuffle the ages randomly and recompute the $\Delta t - R$ relation, using the same bin choices as Figure \ref{fig:agedependency} for the 1--300 Myr clusters.  Figure \ref{fig:randomizeage} shows 50,000 iterations of randomly shuffling the ages and keeping the separations the same.  We see that the act of shuffling the ages causes the correlation to disappear, implying that the observed $\Delta t - R$ relation is statistically significant and is not a random effect.

For each iteration, we also determine the slope, where we only fit cluster pairs up to 500 pc to remove any uncertainty in the location of $R_{\rm max}$.  In the 50,000 trials, there was never a random trial with a slope in agreement within 1$\sigma$ of the slope obtained for the real data of NGC 628 (0.33 $\pm$ 0.07).  This provides further evidence of a physical effect where clusters tend to form coevally, while distant pairs tend to form at different times; the effect is not the result of chance alignment or biases in the survey selection.

\subsubsection{Randomizing the Positions}\label{sec:randompos}
The second test we perform to check whether the age distribution is driving the observed behavior is assigning a random spatial position within the footprint of NGC 628 to each of the fitted cluster ages.  Figure \ref{fig:randomizeage} shows the $\Delta t - R$ relation results for 15,000 iterations of randomizing the positions of the star clusters.  This test of randomizing the positions is more involved than simply shuffling the ages as we have to ensure that enough cluster pairs are within the smallest separation bin in Figure \ref{fig:randomizeage}.  Because we use the actual cluster positions when we randomize the ages in Section \ref{sec:ageshuffle}, we did not have this problem in that first test.

The observational bins in NGC 628 contains 58 cluster pairs within the smallest separation bin centered at 10 pc, and 86 cluster pairs with separations in the 20 pc bin.  We place the requirement of a minimum of 25 pairs in the least populated bin, though the results do not change if we lower the threshold.  In an initial run of 50,000 iterations, using the same bin sizes as Figure \ref{fig:agedependency}, only two trials populated the minimum of 25 cluster pairs in the 10 pc bin, far short from the 58 cluster pairs that are observed in the true cluster positions of NGC 628 in the same bin.  We already infer from this that the positions of the clusters are not random and the $\Delta t - R$ relation cannot be the result of a statistical effect and the probability of a chance alignment is nearly negligible.  Indeed, the clusters of NGC 628 do show a very clustered distribution, especially at young ages, compared to randomized populations \citep{grasha15}.  

To ease computational time and deal with the dearth of close cluster pairs in the randomized positions, we decrease the number of bins by half, increasing the average separation of the smallest bin centered at 10 pc to around 20 pc.  We still require the minimum of 25 clusters in the smallest bin.  On average, only 5\% of the iterations have enough close clusters such that 25 points populated the smallest bin at 20 pc, i.e., in order to obtain 15,000 successful runs with enough clusters populating the right panels of Figure \ref{fig:randomizeage}, we performed over 300,000 trials.  This is a significant result as well, even after requiring a minimum of 25 points with separations around 20 pc, still substantially less than what is observed in real clusters of NGC 628 with 86 cluster pairs at these separations, the cluster positions are not driven by randomized processes.  

With similar findings with the shuffled age test, we find that the correlation disappears when the positions are randomized, suggesting that the $\Delta t - R$ correlation is not a statistical effect.  This is further reinforced by the lack of multiple cluster pairs at short distances in the randomized data compared to the presence of a significant number of close, young clusters pairs observed between real clusters.  For 15,000 successful trials of randomizing the positions, the probability of a random trial having a slope in agreement within 1$\sigma$ of the value obtained for the real $\Delta t - R$ correlation of NGC 628 (0.33 $\pm$ 0.07) is 0.02\% (3 out of 15,000 trials).  The cluster positions, combined with their ages, are indeed imprinted with information on the physical mechanism that drives cluster formation and the $\Delta t - R$ correlation is statistically robust.

\subsubsection{Unbinned Results}
\begin{figure*}
\epsscale{1.1}
\plottwo{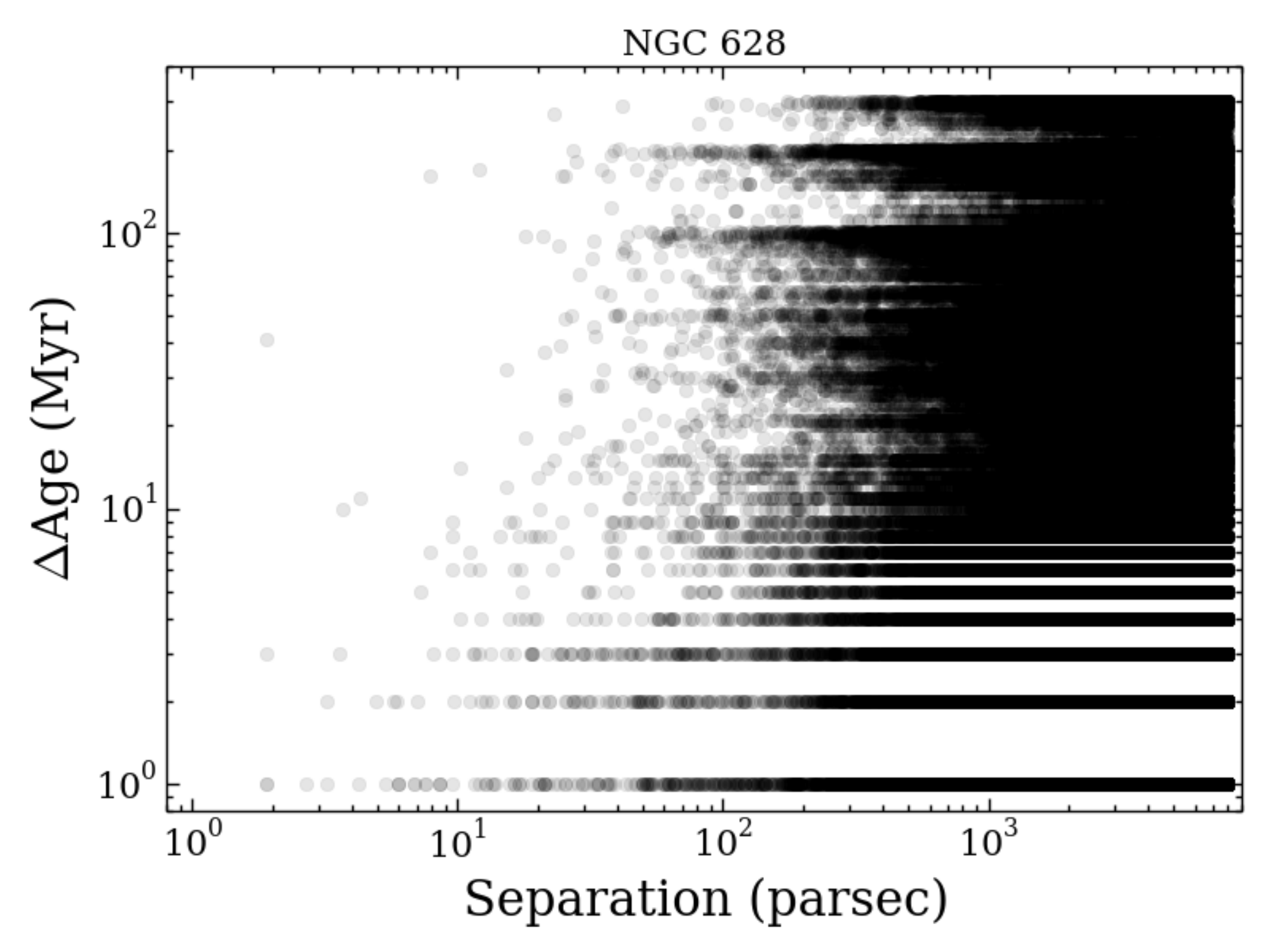}{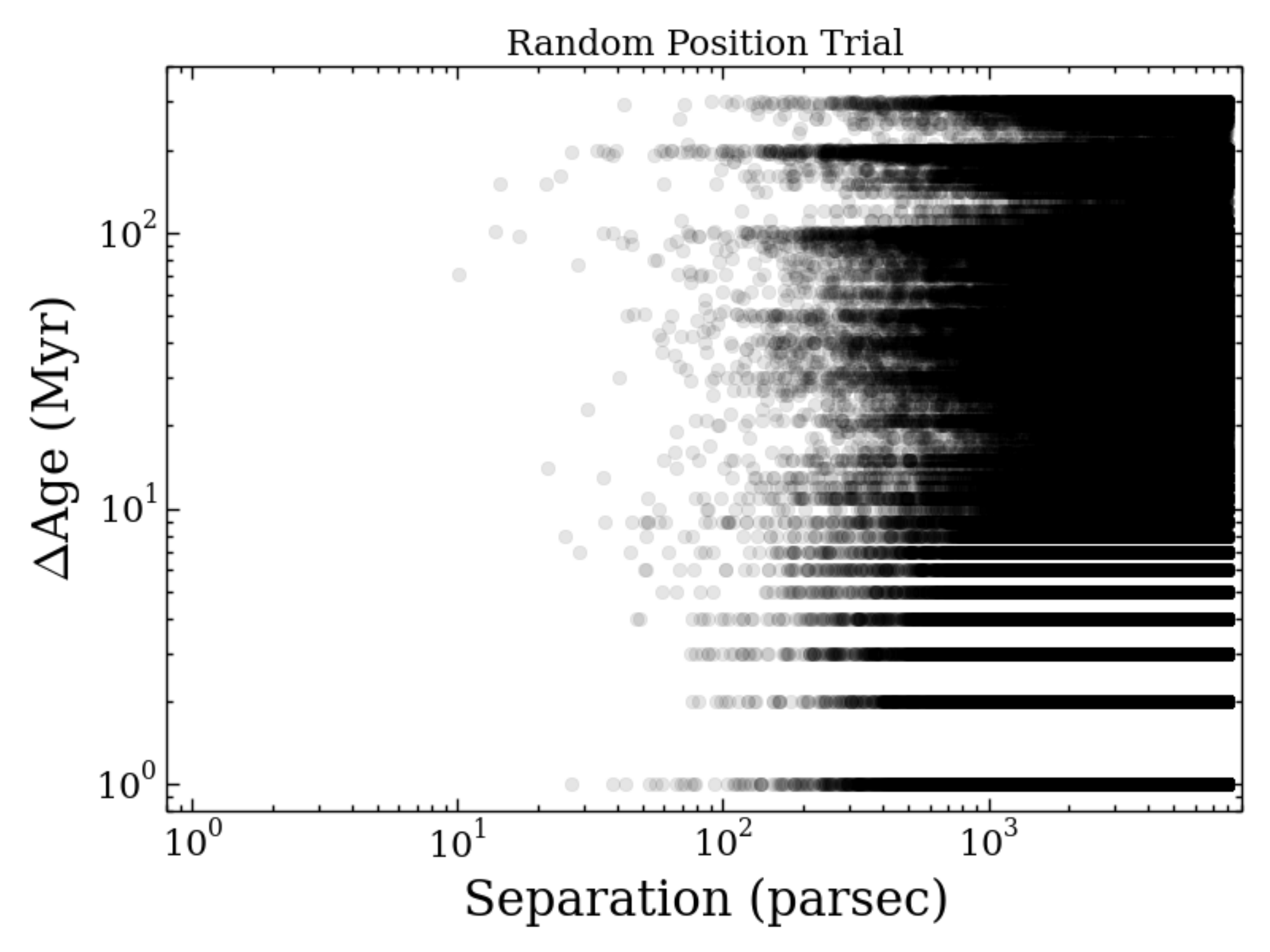}
\caption{
The age difference between cluster pairs as a function of separation between the cluster pairs without any binning for NGC 628 and the random cluster position trial in Figure \ref{fig:randomizeage}.  There is an excess of clusters at smaller separations for smaller age differences in the actual data, which is not recovered in the simulations. 
\label{fig:agespreadtrue}}
\end{figure*}

As briefly discussion in Section \ref{sec:agedependency}, the scatter in the intrinsic $\Delta t - R$ across all separation lengths is considerable and the clusters will gather in rows along the y-axis owing to the discrete ages resulting from SED fitting.  Figure \ref{fig:agespreadtrue} shows the unbinned distribution for all the cluster pairs in NGC 628.   We compare it to the one of the 15,000 trials where the clusters are assigned randomized positions (top right panel of Figure \ref{fig:randomizeage}).  The cluster ages are the same in both cases, only the positions are changing between the two panels of Figure \ref{fig:agespreadtrue}.  There is a noticeable increase in younger cluster pairs at small separations of less than a few hundred parsecs in the real data that is not present in the randomized trial, which becomes more apparent when we bin across spatial separations in Figure \ref{fig:randomizeage}.  In the real data, the majority of clusters with separations less than 50 pc are generally less than $\sim$30 Myr old, while in the random trial, there is no trend in age for clusters pair separations less than 50 pc.  Therefore, despite the scatter present, the unbinned data support the inference from the binned data, that the $\Delta t - R$ relation is statistically robust, further supporting our results from the previous section using both randomized ages and randomized positions.

\section{Discussion}\label{sec:discussion}
\citet{larson81} was the first to establish that turbulence is responsible for the observed velocity dispersion and size correlation of GMCs, where the line widths increase as a power of their radius:  $\sigma \propto R^{\beta}$ with $\beta = 0.38$.  This relation indicates that turbulence is faster in larger regions.  The slope for the size--velocity dispersion relationship is now believed to be slightly steeper: $\beta \propto 0.5$ \citep[e.g., ][]{solomon87,rosolowsky08,rice16}.  Our results show that the duration of star formation within star-forming regions in eight galaxies is $\Delta t \propto R^{[0.25,0.6]}$, suggesting that the $\Delta t - R$ cluster relation is consistent with resulting from turbulence.  For a hierarchical ISM arising from turbulence, the velocity difference $\Delta v$ between points separated in space by $\Delta x$ varies as $\Delta v \propto \Delta x^\beta$ and the crossing time $\Delta t$ over that distance varies as $\Delta t \propto \Delta x / \Delta v \propto \Delta x^{1-\beta}$.  Thus, our slope $\alpha$ equals $1-\beta$ for slope $\beta$ in the size-linewidth relation of interstellar turbulence.  For a turbulent model, $\beta\sim0.5$ \citep{rice16}, results in $\alpha\sim0.5$.  This value of $\alpha$ is consistent with observations in the LMC \citep{elmegreen96,efremov98}, the MW \citep{delafuentemarcos09}, and with our sample.  Furthermore, correctly accounting for the effects of cluster evolution brings the observed slopes closer to the expected value of 0.5 from turbulence \citep[see Section \ref{sec:dissolution} and][]{delafuentemarcos09}. 

The change from a power-law distribution to a flat distribution in the $\Delta t - R$ cluster relation may describe the transition from a scale-free turbulent motion at small-scales to uniform large-scale galactic dynamics, occurring around 0.5--1~kpc \citep{elmegreenetal96,sanchez10,dutta13}.  It is likely related to the maximum size of a coherent star-forming region, which may be the turbulent Jeans length, expected from self-gravity on the largest scale \citep{elmegreen83}, or the length given by galaxy rotation \citep{escala08}.  
The separation sizes are also similar to the maximum correlated size of cluster complexes as probed using the two-point correlation function in the same galaxies \citep{grasha15,grasha17}.  

While the age and separations of the youngest clusters can be well-described with a turbulent model, a better understanding on the effects of cluster evolution and survival can help improve our understanding of turbulent-driven star formation and unravel whether the observed deviations of the slopes from the theoretical 0.5 value are the results of other evolutionary or environmental effects.  The maximum size of a turbulent cell appears to be limited by shear: we find a tight relation between the maximum turbulent velocity and the average velocity difference due to shear at the size of the largest correlated star-forming region (Figure \ref{fig:shear}).  

The combination of both the galaxy size and the shear determines the average age of star-forming regions in the $\Delta t - R$ correlation, varying by a factor of five across the sample.  Larger galaxies exhibit larger star-forming complexes, impacting the size of $R_{\rm max}$, and within each complex, the age difference of a sub-region scales with roughly the square root of the size of the sub-region (Figure \ref{fig:agespread}).  Shear limits the average ages of the regions by limiting the ratio of the maximum size to the average age.  The duration of star formation within a region is proportional to a few crossing times and the recovered velocities are similar to the turbulence motions of the ISM \citep[$\sim 10$~\kms;][]{heiles03}.  

Clusters formed together in groups that exhibit similar ages is an expected outcome of hierarchical star formation model \citep{bhatia88,dieball02,desilva15}.  The relation between time and distance we find implies that young clusters born together in close pairs (or groupings) do indeed show similar ages, providing further support that star formation, as traced by these young stellar clusters, is organized in a hierarchical manner \citep[e.g., ][]{hopkins13}.  The $\Delta t - R$ relation is statistically robust and significant, the correlation disappearing when both the ages and positions are randomized within each galaxy (Section \ref{sec:tests}).  Star formation is not a random process: both the positions and ages are consistent with being driven by turbulence and it is unlikely to have a random process give rise to the large number of close-age cluster pairs observed in local galaxies.

\section{Summary and Conclusion}\label{sec:summary}
In this work, we investigate the relation between age difference and the separation of young ($<$300 Myr) cluster pairs in eight local galaxies to investigate whether a correlation exists between the duration of star formation and the size of the star-forming region.  
Our main results can be summarized as follows: 
\begin{enumerate}
\item Clusters that are born closer to each other exhibit smaller age differences compared to clusters that are born further apart.  This time-distance correlation among young clusters implies that on average younger star-forming structures are less extended than older regions.  
\item The average age difference between pairs of clusters increases with their separation as the 0.25 to 0.6 power.  The duration of star formation and the size of the star forming regions are consistent with expectations from turbulence: larger regions can form stars over longer timescales as the duration of star formation scales with the square root of the size over longer timescales as an effect of turbulence cascading down to smaller size regions.  The slopes we recover tend to be closer to 0.5 if only clusters younger than $\sim$100 Myr are considered, indicating that the hierarchical distribution, inherited by a turbulent ISM, dissipates with time. 
\item The power-law relation reaches a maximum size where the age difference between pairs becomes constant.  The maximum size increases with galaxy size, and the ratio of the maximum size to the average age difference -- values that are similar to the turbulent speed of the ISM -- increases with galaxy shear.  These results imply that star formation proceeds hierarchically in giant star complexes with a duration on small scales that is proportional to the local turbulent crossing time.  The complexes are then dispersed by shear.
\item For the spiral galaxies in our sample, the maximum velocity marginally correlates with the galaxy's size; weaker correlations are found with the galaxy's SFR and morphology.  The dwarf galaxy in our sample does not fit the trend of the spirals. 
\end{enumerate}

At small scales, turbulence appears to be a primary candidate driver of the spatial and temporal scales for star formation processes although other potential drivers, like self-gravity \citep{li05} will need to be further investigated.  Future studies and improved simulations over an expanded range of galactic parameters that also consider cluster feedback into the ISM will help provide the necessary insight to the secondary processes that impact the overall organization of star formation across a wide range of scales.

\acknowledgements
We are indebted to an anonymous referee for their careful reading of this manuscript and for providing comments that have improved the scientific outcome and quality of the paper.
We also appreciate the very useful discussions and comments on this work by R. Klessen, D. Kruijssen, and B. Gaches. 
Based on observations made with the NASA/ESA Hubble Space Telescope, obtained at the Space Telescope Science Institute, which is operated by the Association of Universities for Research in Astronomy, under NASA Contract NAS 5--26555.  These observations are associated with Program 13364 (LEGUS).  Support for Program 13364 was provided by NASA through a grant from the Space Telescope Science Institute.  This research has made use of the NASA/IPAC Extragalactic Database (NED) which is operated by the Jet Propulsion Laboratory, California Institute of Technology, under contract with NASA.



\end{document}